\documentclass[sigplan, 9pt]{acmart}

\AtBeginDocument{%
  \providecommand\BibTeX{{%
    \normalfont B\kern-0.5em{\scshape i\kern-0.25em b}\kern-0.8em\TeX}}}

\setcopyright{acmcopyright}
\copyrightyear{2020}
\acmYear{2020}
\acmDOI{}

\acmConference[ICCAD '20]{}{}{}
\acmBooktitle{}
\acmPrice{}
\acmISBN{}
\acmSubmissionID{75}

\usepackage{subfiles}
\usepackage{listings}
\usepackage{graphicx}
\usepackage{subcaption}
\usepackage{multirow}
\usepackage{mathtools}
\usepackage{tikz}
\usepackage{pgfplots}
\usetikzlibrary{positioning, decorations.pathreplacing, calc, intersections, pgfplots.groupplots}

\begin{document}

\title{Optimal Layout Synthesis for Quantum Computing}

\author{Bochen Tan}
\affiliation{
  \institution{University of California, Los Angeles}
}
\email{bctan@cs.ucla.edu}

\author{Jason Cong}
\affiliation{
  \institution{University of California, Los Angeles}
}
\email{cong@cs.ucla.edu}

\begin{abstract}
Recent years have witnessed the fast development of quantum computing.
Researchers around the world are eager to run larger and larger quantum algorithms that promise speedups impossible to any classical algorithm.
However, the available quantum computers are still volatile and error-prone.
Thus, layout synthesis, which transforms quantum programs to meet these hardware limitations, is a crucial step in the realization of quantum computing.
In this paper, we present two synthesizers, one optimal and one approximate but nearly optimal.
Although a few optimal approaches to this problem have been published, our optimal synthesizer explores a larger solution space, thus is optimal in a stronger sense.
In addition, it reduces time and space complexity exponentially compared to some leading optimal approaches.
The key to this success is a more efficient spacetime-based variable encoding of the layout synthesis problem as a mathematical programming problem.
By slightly changing our formulation, we arrive at an approximate synthesizer that is even more efficient and outperforms some leading heuristic approaches, in terms of additional gate cost, by up to 100\%, and also fidelity by up to 10x on a comprehensive set of benchmark programs and architectures.
For a specific family of quantum programs named QAOA, which is deemed to be a promising application for near-term quantum computers, we further adjust the approximate synthesizer by taking commutation into consideration, achieving up to 75\% reduction in depth and up to 65\% reduction in additional cost compared to the tool used in a leading QAOA study.
\end{abstract}



\maketitle

\section{Introduction}
For years, quantum algorithms have been shown theoretically to hold significant advantages over classical algorithms on some algebraic problems like factoring large numbers~\cite{sfcs94-shor-algorithms}, quantum chemistry~\cite{science05-aspuru-guzik-dutoi-love-head-gordon-molecular-energies}, machine learning~\cite{nature17-biamonte-wittek-pancotti-rebentrost-wiebe-lloyd-qml}, and so on.
Recently, the first experimental proof of quantum computational advantage on certain sampling problems was published~\cite{nature19-google-supremacy}.

To implement a quantum algorithm on an actual quantum computer, \textit{logic synthesis} and \textit{layout synthesis} for quantum computing (QC) are inevitable.
A classical algorithm is a series of operations on a set of bits; likewise, a quantum algorithm is a series of operations, i.e., \textit{quantum gates}, on a set of quantum bits, i.e., \textit{qubits}.
A realistic \textit{quantum computer architecture} only supports gates in its \textit{gate library}.
To implement a quantum algorithm, the original quantum gates have to be translated into gates within this library.
This process is called logic synthesis for quantum computing.
It has been shown that all quantum algorithms can be decomposed to quantum gates acting on a single qubit or two qubits ~\cite{book10-nielsen-chuang}, so usually the output of logic synthesis is a list of single-qubit gates and two-qubit gates.

After logic synthesis, the qubits and quantum gates are still purely logical.
Next, a mapping from such \textit{logical qubits} to the \textit{physical qubits} on QC architecture and a schedule which decides when every quantum gate is executed have to be provided.
(Note that our notion of logical qubit is different from that in the quantum error correction.)
The task of deriving such information is called layout synthesis for quantum computing (LSQC)~\cite{tc20-tan-cong-optimality}.
Other names have also been used to refer to this task, e.g., quantum circuit compilation~\cite{dac20-alam-ash-saki-ghosh-qaoa-compilation,tqc19-childs-shoute-unsal-transformation}, transpilation (in Qiskit\footnote{\url{https://qiskit.org/}}), placement~\cite{tcad08-maslov-falconer-mosca-placement,aspdac14-shafaei-saeedi-pedram-2014-placement-2d}, mapping~\cite{asplos19-li-ding-xie-sabre,dac19-wille-burgholzer-zulehner-minimal,date18-zulehner-paler-wille-efficient,isca19-murali-linke-martonosi-abhari-nguyen-triq,iccad19-bhattacharjee-saki-alam-chattopadhyay-ghosh-muqut}, qubit allocation~\cite{cgo18-siraichi-santos-collange-pereira-allocation}, and routing~\cite{arxiv2003-sivarajah-dilkes-cowtan-simmons-edgington-duncan-tket}.
We choose to use the term `layout synthesis' as it involves both placement and scheduling.
There is an important issue to resolve in LSQC.
Every two-qubit gate requires a connection between its two qubits.
An algorithm designer may assume the logical qubits to have all-to-all connectivity, but this is not the case for the physical qubits in many QC architectures, e.g., the superconducting QC systems at Google~\cite{nature19-google-supremacy}.
As a result, every QC architecture comes with a \textit{coupling graph} specifying which physical qubit pairs are connected.
If the pair of logical qubits of a two-qubit gate is mapped to a disconnected physical qubit pair on the coupling graph, this gate will not be executable.
Luckily, there is a special type of gate, SWAP, that can exchange the mapping of two logical qubits.
Thus, with a series of SWAP gates, two logical qubits can reach two connected physical qubits, but this comes at some additional cost.
The decisions of when and where to apply these SWAP gates are made in LSQC.

A recent study~\cite{tc20-tan-cong-optimality} revealed large optimality gaps of several leading LSQC tools with a set of benchmarks named QUEKO that have known optimal solutions.
Therefore, despite the NP-completeness of the problem~\cite{cgo18-siraichi-santos-collange-pereira-allocation, tcad08-maslov-falconer-mosca-placement,tc20-tan-cong-optimality}, optimal approaches to LSQC are still worthy of research because they provide the baseline of measuring the available LSQC solutions.
There have been a few previous works on exact or optimal LSQC~\cite{aspdac14-wille-lye-dreschsler-optimal, aspdac14-shafaei-saeedi-pedram-2014-placement-2d, dac19-wille-burgholzer-zulehner-minimal, iccad19-bhattacharjee-saki-alam-chattopadhyay-ghosh-muqut, cgo18-siraichi-santos-collange-pereira-allocation}.
However, they either just focus on specific coupling graphs, or has some implicit and unnecessary constraints.
As for the heuristic approaches~\cite{date18-zulehner-paler-wille-efficient, tcad08-maslov-falconer-mosca-placement, cgo18-siraichi-santos-collange-pereira-allocation, tqc19-childs-shoute-unsal-transformation, asplos19-tannu-qureshi-variability, asplos19-li-ding-xie-sabre, isca19-murali-linke-martonosi-abhari-nguyen-triq, arxiv2003-sivarajah-dilkes-cowtan-simmons-edgington-duncan-tket}, the optimality gaps suggest that they still have a lot of room for improvement. 

In this paper, we present two layout synthesizers and compare them with leading related works on a comprehensive set of benchmark quantum programs and architectures.
The optimal layout synthesizer for quantum computing (OLSQ) formulates LSQC as a satisfiability modulo theories optimization problem (SMT).
This problem is then passed to Z3 SMT solver~\cite{book08-de-moura-bjorner-z3} which guarantees optimal solutions.
Compared to Wille et al.~\cite{dac19-wille-burgholzer-zulehner-minimal}, a leading optimal approach which also formulates LSQC to SMT, we provide exponential reduction on the number of variables, resulting in orders-of-magnitude reductions in runtime and the memory usage.
The key to this improvement is to represent the solution space with mapping variables, not with mapping transformation variables as in~\cite{dac19-wille-burgholzer-zulehner-minimal,cgo18-siraichi-santos-collange-pereira-allocation}.
Also, by relaxing the gate placement under the \textit{dependency} constraints, OLSQ explores a larger solution space than other gate-by-gate~\cite{dac19-wille-burgholzer-zulehner-minimal,cgo18-siraichi-santos-collange-pereira-allocation} and level-by-level~\cite{iccad19-bhattacharjee-saki-alam-chattopadhyay-ghosh-muqut} ``optimal'' approaches, so its results are sometimes even better.
Furthermore, by eliminating redundant mapping variables between \textit{transitions}, i.e., when the mapping changes due to some SWAP gates, we significantly reduce the number of variables and derive a highly scalable approximate synthesizer, transition-based-OLSQ (TB-OLSQ).
Evaluation results show that TB-OLSQ is nearly optimal and is able to increase fidelity by 1.30x in geomean, compared to TriQ~\cite{isca19-murali-linke-martonosi-abhari-nguyen-triq}, a leading academic work, and reduce cost by 69.2\% in geomean, compared to $\text{t}|\text{ket}\rangle$~\cite{arxiv2003-sivarajah-dilkes-cowtan-simmons-edgington-duncan-tket}, a leading industry work.
Finally, given that quantum approximate optimization (QAOA)~\cite{ arxiv2004-google-qaoa, arxiv1905-nasa-qaoa, arxiv1411-farhi-goldstone-gutmann-qaoa, algorithms19-hadfield-wang-ogorman-rieffel-venturelli-biswas-qaoa} is very much in the spotlight due to its feasibility for near-term quantum computers, we study the LSQC of QAOA.
The structure of gates in QAOA makes it an excellent showcase for the power of our transition-based model.
Moreover, with additional prior knowledge on commutation relations in QAOA, we improve TB-OLSQ for QAOA, denoted as QAOA-OLSQ.
Compared to $\text{t}|\text{ket}\rangle$, which was utilized in a leading QAOA experiment study~\cite{arxiv2004-google-qaoa}, QAOA-OLSQ is able to reduce depth by 70.2\% in geomean and cost by 53.8\% in geomean.

\section{Background}

The input to an LSQC problem consists of two parts.
The first part is a \textit{quantum program} or a \textit{quantum circuit} which is specified as a list of quantum gates or operations $g_0g_1...g_{L-1}$ on logical qubits $q_0,q_1,...,q_{M-1}$.
(There are, in total, $L$ gates and $M$ logical qubits.)
Because we assume that logic synthesis is performed prior to LSQC, all of these gates are either single-qubit or two-qubit gates and must be in the gate library of the architecture.
An important single-qubit gates is $X$ gate, which negates the qubit; an important two-qubit gate is $CX$ gate, which does not modify the first qubit but changes the second qubit to the XOR of the first and second qubits.
Until further notice, we use the Qiskit gate library, which contains $X$ and $CX$ mentioned above, and also other single-qubit gates such as $H$, $T$, $T^\dag$, and $S$.
The specific functions of these gates do not concern LSQC; in fact, we only care about on which qubit(s) the gates operate.

\begin{figure}[tb]
    \centering
    \includegraphics[width=0.4\textwidth]{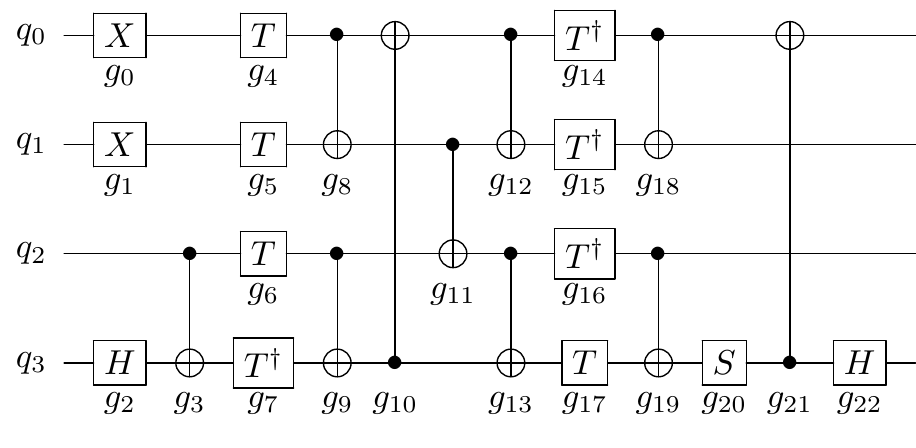}
    \vspace{-1.5\baselineskip}
    \caption{Circuit diagram for quantum adder}
    \label{fig:adder_diagram}
\end{figure}

For example, the beginning of a quantum adder program introduced in~\cite{tcad13-amy-maslov-mosca-roeteller-meetinthemiddle} is shown below.
\begin{verbatim}
OPENQASM 2.0;         // code for quantum adder
include "qelib1.inc"; // gate library
qreg q[4];            // declare 4 logical qubits
x q[0];               // g_0
x q[1];               // g_1
h q[3];               // g_2
cx q[2], q[3];        // g_3
t q[0];               // g_4
...                   ...
\end{verbatim}
In the above program, $g_0$ is an $X$ gate on logical qubit $q_0$; $g_3$ is a $CX$ gate, on $q_2$ and $q_3$.
One way to visualize a quantum program is a \textit{circuit diagram}, e.g., Figure~\ref{fig:adder_diagram}, where each wire stands for a logical qubit, and the gates are placed from left to right on the wires according to the program order.
To make a distinction between single-qubit and two-qubit gates, we separate them into two lists $G_1$ and $G_2$.
For $g_l\in G_1$, we use $g_l.q$ to denote the logical qubit it operates on; for $g_l\in G_2$, we use $g_l.q$ and $g_l.q'$.
For example, $g_0.q=q_0$, $g_5.q=q_1$, $g_3.q=q_2$, $g_3.q'=q_3$.

\small
\normalsize

\begin{figure}[tb]
    \begin{minipage}[b]{0.2\textwidth}
        \includegraphics[width=\linewidth]{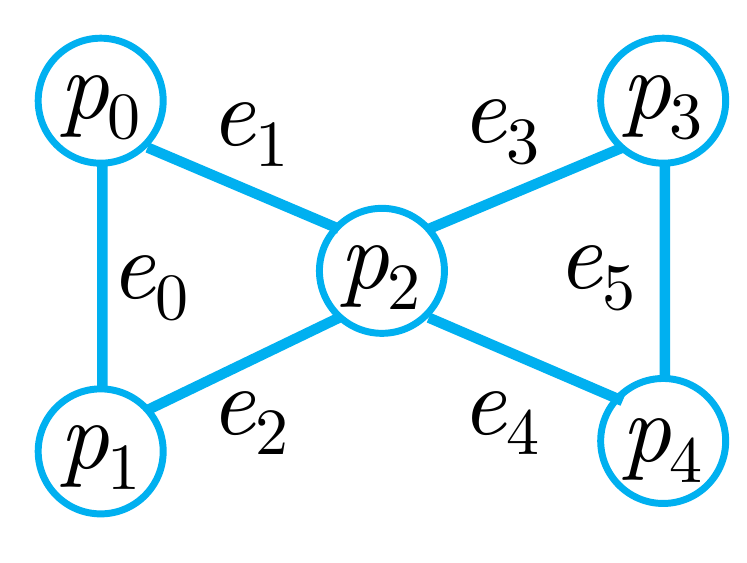}\\
        \vspace{-1.5\baselineskip}
        \subcaption{IBM QX2}
        \label{fig:coupling_graph_ibmqx2}
    \end{minipage}
    \hspace{1em}
    \begin{minipage}[b]{0.2\textwidth}
        \includegraphics[width=\linewidth]{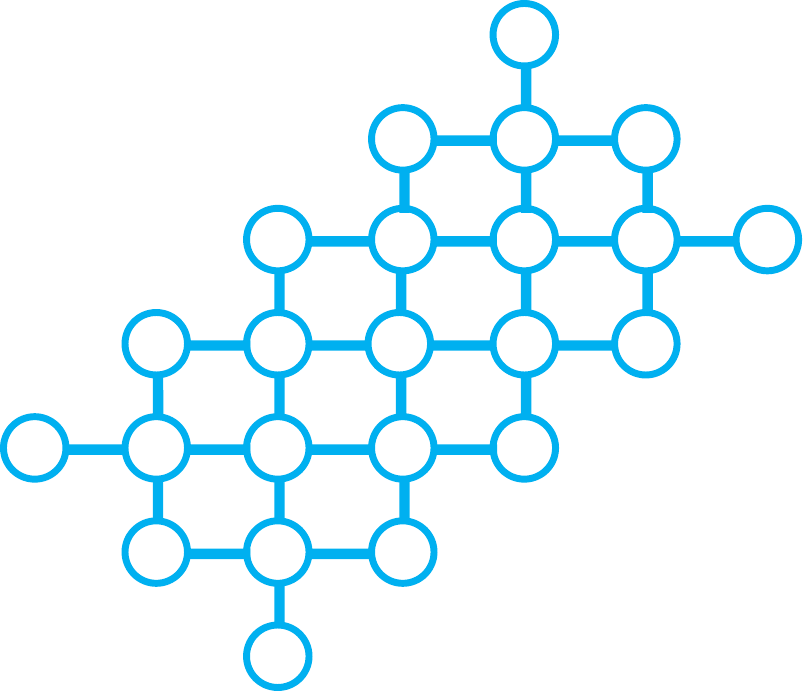}\\
        \vspace{-1.5\baselineskip}
        \subcaption{Google Sycamore (part of)}
        \label{fig:coupling_graph_sycamore23}
    \end{minipage}
    
    \bigskip
    \begin{minipage}[b]{0.18\textwidth}
        \includegraphics[width=\linewidth]{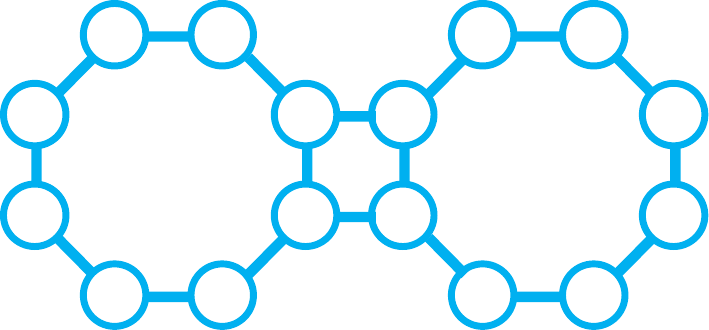}\\
        \vspace{-1.5\baselineskip}
        \subcaption{Rigetti Aspen-4}
        \label{fig:coupling_graph_aspen4}
    \end{minipage}
    \hspace{1em}
    \begin{minipage}[b]{0.22\textwidth}
        \includegraphics[width=\linewidth]{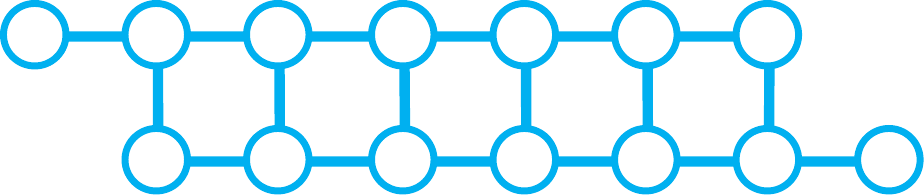}\\
        \vspace{-1.5\baselineskip}
        \subcaption{IBM Melbourne}
        \label{fig:coupling_graph_melbourne}
    \end{minipage}
    \vspace{-1\baselineskip}
    \caption{Coupling Graphs}
    \label{fig:coupling_graph}
\end{figure}

The second part of the input to LSQC problem is a coupling graph $(P,E)$, where $P=\{p_0,p_1,...,p_{N-1}\}$ is the set of physical qubits and $E=\{e_0,e_1,...,e_{K-1}\}$ is the set of (undirected) connections between them.
(There are, in total, $N$ physical qubits and $K$ edges.)
The coupling graphs used in this paper are shown in Figure~\ref{fig:coupling_graph}.
We denote an edge as $(e_k.p, e_k.p')$, e.g., in Figure~\ref{fig:coupling_graph_ibmqx2}, $e_1.p=p_0$ and $e_1.p'=p_2$.
In addition, fidelity information can be provided as three functions: $f_0:P\to [0,1]$ for measurements, $f_1:P\to [0,1]$ for single-qubit gates, and $f_2:E\to [0,1]$ for two-qubit gates.
(Under current technology constraints, quantum computer providers usually only offer one type of entangling two-qubit gate.
We are referring the fidelity for this type of gate.)

\begin{figure}[tb]
    \centering
    \includegraphics[width=\columnwidth]{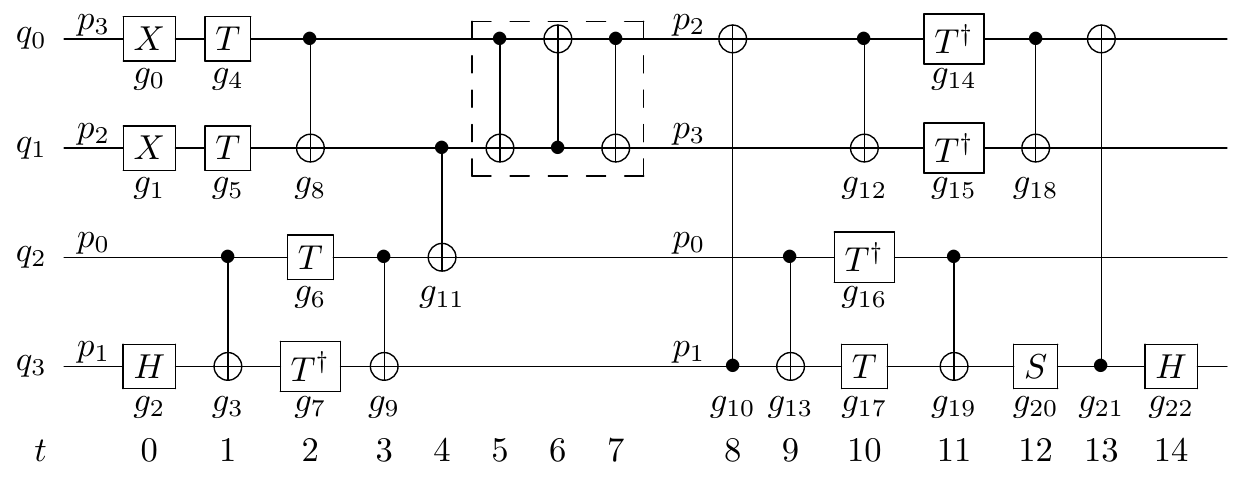}
    \vspace{-2\baselineskip}
    \caption{Result of LSQC for quantum adder}
    \label{fig:adder_ibmqx2}
\end{figure}

The output of LSQC is the \textit{spacetime coordinates} $(t_l, x_l)$ for all the input gates and the SWAP gates inserted, and a final mapping $\pi:Q\to P$ providing which physical qubit to measure for each logical qubit.
We show the LSQC result of the quantum adder (Figure~\ref{fig:adder_diagram}) on IBM QX2 (Figure~\ref{fig:coupling_graph_ibmqx2}) in Figure~\ref{fig:adder_ibmqx2}.
In this diagram, gates are aligned according to their time coordinates, e.g., $t_0=t_1=t_2=0$ and $t_{10}=8$; the space coordinates of single-qubit gates can be read off from the mapping displayed above the wires, e.g., $q_0$ is mapped to $p_3$ at time $0$, so $x_0=p_3$; the space coordinates of two-qubit gates can be deduced from the mapping, e.g., at time $8$, $q_3$ and $q_0$ are mapped correspondingly to $p_1$ and $p_2$, so $x_{10}=e_2$ because $e_2$ connects $p_1$ and $p_2$.
The mapping remains the same as the previous time slot if it is not displayed, e.g., $q_1$ is still mapped to $p_2$ at time $1$, so $x_5=p_2$.
Thus, the final mapping is just the last mapping displayed, $\pi(q_0)=p_2$, $\pi(q_1)=p_3$, $\pi(q_2)=p_0$, and $\pi(q_3)=p_1$.
A SWAP gate consisting of three $CX$ gates is inserted on $e_3$ between $p_2$ and $p_3$.
We use the last time slot each SWAP gate takes as its time coordinate (in this case, $7$).

The inserted SWAP gate in this example is absolutely necessary.
The quantum program has two-qubit gates between $q_0$ and $q_1$ ($g_8$, $g_{12}$, and $g_{18}$), $q_1$ and $q_2$ ($g_{11}$), $q_2$ and $q_3$ ($g_3$, $g_9$, $g_{13}$, and $g_{19}$), and finally $q_3$ and $q_0$ ($g_{10}$ and $g_{21}$).
This means, without any SWAP gates, the logical qubits must be mapped to a set of physical qubits connected like a square.
However, the coupling graph, Figure~\ref{fig:coupling_graph_ibmqx2}, does not contain such a structure.

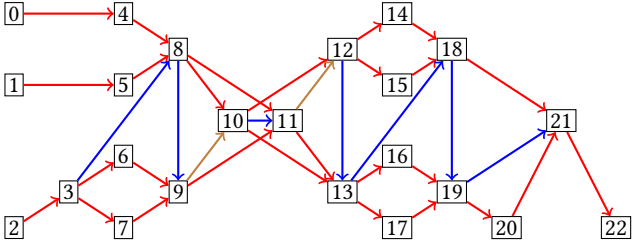
\begin{figure}[tb]
\centering
    \begin{tikzpicture}[x=2.3em, y=3em, inner sep=0.15em]
        \node [draw] (g0) at (0,3) {$0$};
        \node [draw] (g1) at (0,2) {$1$};
        \node [draw] (g2) at (0,0) {$2$};
        \node [draw] (g3) at (1,0.5) {$3$};
        \node [draw] (g4) at (2,3) {$4$};
        \node [draw] (g5) at (2,2) {$5$};
        \node [draw] (g6) at (2,1) {$6$};
        \node [draw] (g7) at (2,0) {$7$};
        \node [draw] (g8) at (3,2.5) {$8$};
        \node [draw] (g9) at (3,0.5) {$9$};
        \node [draw] (g10) at (4,1.5) {${10}$};
        \node [draw] (g11) at (5,1.5) {${11}$};
        \node [draw] (g12) at (6,2.5) {${12}$};
        \node [draw] (g13) at (6,0.5) {${13}$};
        \node [draw] (g14) at (7,3) {${14}$};
        \node [draw] (g15) at (7,2) {${15}$};
        \node [draw] (g16) at (7,1) {${16}$};
        \node [draw] (g17) at (7,0) {${17}$};
        \node [draw] (g18) at (8,2.5) {${18}$};
        \node [draw] (g19) at (8,0.5) {${19}$};
        \node [draw] (g20) at (9,0) {${20}$};
        \node [draw] (g21) at (10,1.5) {${21}$};
        \node [draw] (g22) at (11,0) {${22}$};
        
        \path[draw, ->, thick, red] (g0) edge (g4);
        \path[draw, ->, thick, red] (g4) edge (g8);
        \path[draw, ->, thick, red] (g8) edge (g10);
        \path[draw, ->, thick, red] (g10) edge (g12);
        \path[draw, ->, thick, red] (g12) edge (g14);
        \path[draw, ->, thick, red] (g14) edge (g18);
        \path[draw, ->, thick, red] (g18) edge (g21);

        \path[draw, ->, thick, red] (g1) edge (g5);
        \path[draw, ->, thick, red] (g5) edge (g8);
        \path[draw, ->, thick, red] (g8) edge (g11);
        \path[draw, ->, thick, brown] (g11) edge (g12);
        \path[draw, ->, thick, red] (g12) edge (g15);
        \path[draw, ->, thick, red] (g15) edge (g18);

        \path[draw, ->, thick, red] (g3) edge (g6);
        \path[draw, ->, thick, red] (g6) edge (g9);
        \path[draw, ->, thick, red] (g9) edge (g11);
        \path[draw, ->, thick, red] (g11) edge (g13);
        \path[draw, ->, thick, red] (g13) edge (g16);
        \path[draw, ->, thick, red] (g16) edge (g19);

        \path[draw, ->, thick, red] (g2) edge (g3);
        \path[draw, ->, thick, red] (g3) edge (g7);
        \path[draw, ->, thick, red] (g7) edge (g9);
        \path[draw, ->, thick, brown] (g9) edge (g10);
        \path[draw, ->, thick, red] (g10) edge (g13);
        \path[draw, ->, thick, red] (g13) edge (g17);
        \path[draw, ->, thick, red] (g17) edge (g19);
        \path[draw, ->, thick, red] (g19) edge (g20);
        \path[draw, ->, thick, red] (g20) edge (g21);
        \path[draw, ->, thick, red] (g21) edge (g22);
        
        \path[draw, ->, thick, blue] (g3) edge (g8);
        \path[draw, ->, thick, blue] (g8) edge (g9);
        \path[draw, ->, thick, blue] (g10) edge (g11);
        \path[draw, ->, thick, blue] (g12) edge (g13);
        \path[draw, ->, thick, blue] (g13) edge (g18);
        \path[draw, ->, thick, blue] (g18) edge (g19);
        \path[draw, ->, thick, blue] (g19) edge (g21);
        
    \end{tikzpicture}
    \vspace{-0.5\baselineskip}
    \caption{Immediate dependencies in the quantum adder (Red arrows are used in OLSQ. Green arrows are those imposed implicitly by the approach of Wille et al.'s. Brown means intersection.)}
    \label{fig:dependency}
\end{figure}

We stress a few constraints for LSQC. 
\textbf{1)} We assume any kind of gate cancellation/optimization has been performed prior to LSQC, so all the input gates should be executed.
\textbf{2)} The two-qubit gates should be mapped to valid edges on the coupling graph.
\textbf{3)} Avoid \textit{collisions}: if two gates $g_l$ and $g_{l'}$ act on the same qubit, then they cannot be executed at the same time, i.e., $t_l\neq t_{l'}$. For example, $t_5\neq t_8$ because both $g_5$ and $g_8$ act on $q_1$.
\textbf{4)} Respect \textit{dependencies}: if we have a dependency between $g_l$ and $g_{l'}$, $l<l'$, then their relative order should not change, i.e., $t_l<t_{l'}$.
If there are no gates between these two gates, it is called an \textit{immediate dependency}.
However, deriving the dependencies in a generic quantum program turns out to be a nontrivial task.
By default, we treat all the collision gate pairs above as dependencies.
This is a common practice, e.g., also in Murali et al.~\cite{isca19-murali-linke-martonosi-abhari-nguyen-triq}, and also guarantees correctness.
Immediate dependencies of the quantum adder is shown as red and brown arrows in Figure~\ref{fig:dependency}.
With more knowledge on dependencies, we can improve the LSQC solutions, as we shall see in Section~\ref{sec:qaoa-olsq}.

There can be multiple objectives for LSQC:
\textbf{1)} \textit{Depth:} the maximal time coordinate of all gates. 
Due to the limitation of current QC technology, physical qubits can only function well up to a short `lifetime'. 
Minimizing depth is thus very important.
\textbf{2)} \textit{SWAP Cost or gate count:} the total number of additional SWAP gates.
\textbf{3)} \textit{Fidelity:} the product of fidelity of all the single-qubit gates, two-qubit gates, and measurements.
The current quantum computers are error-prone, so maximal fidelity is desired.

\section{Related Works}

\textit{Exact Solutions:} There have been multiple previous exact or optimal approaches related to LSQC. 
Shafaei et al.~\cite{dac13-shafaei-saeedi-pedram-linear} first finds a qubit mapping on a 2D grid that minimizes expected distances between qubit pairs required by two-qubit gates.
Then, if some two-qubit gate still acts on non-adjacent qubits, the xy routing algorithm is applied to move them together.
Wille et al.~\cite{aspdac14-wille-lye-dreschsler-optimal, dac19-wille-burgholzer-zulehner-minimal} use pseudo-Boolean optimizer and SMT solver to minimize additional gate cost; Siraichi et al.~\cite{cgo18-siraichi-santos-collange-pereira-allocation} provide a dynamic programming approach to this task.
They split the quantum program into individual gates and consider changes to the mapping at every interval.
Instead of individual gates, Bhattacharjee et al.~\cite{iccad19-bhattacharjee-saki-alam-chattopadhyay-ghosh-muqut} splits the program into `levels' of gates that can be executed in parallel.
Afterwards, it optimizes depth with integer linear programming to decide the initial mapping and adjustments of mapping between the levels.

We find that either the gate-by-gate or the level-by-level arrangement imposes some additional constraints than arranging the gates with dependency. 
We shall illustrate this point with the aforementioned LSQC instance.
In this case, Wille et al. return a solution with two SWAP gates, but one SWAP would be enough as shown in Figure~\ref{fig:adder_ibmqx2}.
This is because when arranging gate-by-gate, there is an implicit dependency between each two-qubit gate and the one after it, which are shown as blue and brown arrows in Figure~\ref{fig:dependency}.
Specifically, there is a dependency from $g_{10}$ to $g_{11}$, which means that in their result $t_{10}<t_{11}$, but in the optimal solution, Figure~\ref{fig:adder_ibmqx2}, $t_{10}>t_{11}$.
In fact, When we manually impose an additional dependency from $g_{10}$ to $g_{11}$ onto OLSQ-SWAP (to be specified in the next section), it gives a solution with two SWAP gates as well.
In essence, the gate list is only one of the topological orderings of the dependency graph like Figure~\ref{fig:dependency}.
If we use a gate-by-gate arrangement with the input gate list order, the result certainly respects all the dependencies, but there are some additional arbitrarities that may lead to sub-optimality.
OLSQ actually reconstructs the dependency graph from the gate list and thus effectively explores \textit{all} topological orderings.
As for arranging level-by-level, in the LSQC instance above, suppose we assign $g_8$ and $g_9$ as the first level, $g_{10}$ and $g_{11}$ as the second level, $g_{12}$ and $g_{13}$ as the third level.
Then there must be at least one SWAP inserted between level one and two; otherwise there exists a mapping that can satisfy all the two-qubit connections required by $g_8$, $g_9$, $g_{10}$, and $g_{11}$.
This is to say the coupling graph contains a square-like structure, which is impossible in Figure~\ref{fig:coupling_graph_ibmqx2}.
Similarly, there must be at least one SWAP inserted between level two and three.
Thus, at least two SWAP gates are required, which is sub-optimal compared to Figure~\ref{fig:adder_ibmqx2}.

\textit{Heuristic Solutions:} Maslov et al.~\cite{tcad08-maslov-falconer-mosca-placement} publish one of the earliest work in LSQC and propose a heuristic algorithm that recursively cuts the coupling graph to derive SWAP gates to adjust the mapping.
Siraichi et al.~\cite{cgo18-siraichi-santos-collange-pereira-allocation} find the initial mapping by matching out-degrees and search for SWAP gates to apply based on a heuristic distance between qubits.
Zulehner et al.~\cite{date18-zulehner-paler-wille-efficient} partition the program and employs A* search with a lookahead scheme to find the SWAP gates that transform the current mapping to the next.
Li et al.~\cite{asplos19-li-ding-xie-sabre} apply bi-directional search to minimize additional gate cost.
Childs et al.~\cite{tqc19-childs-shoute-unsal-transformation} extend existing approaches to the token swapping problem in order for SWAP insertion.
Tannu\&Qureshi~\cite{asplos19-tannu-qureshi-variability} observe the variation in gate fidelity and thus propose taking fidelity into the LSQC problem.
Murali et al.~\cite{isca19-murali-linke-martonosi-abhari-nguyen-triq} present an end-to-end LSQC flow for fidelity optimization.
An SMT solver is applied to derive the initial mapping.
For SWAP insertion, they construct a reliability matrix of two-qubit gates, which considers the SWAP gates required along the way, and use this matrix to route qubits on the coupling graph.
In $\text{t}|\text{ket}\rangle$~\cite{arxiv2003-sivarajah-dilkes-cowtan-simmons-edgington-duncan-tket}, when some two-qubit gate is not executable, the algorithm keeps the relevant qubit permutations as candidates and treats the gates ahead to evaluate them.
If more than one candidate receives the highest score before the next non-executable gate, the process goes into a new level until one candidate wins.
Recently, Alam et al.~\cite{dac20-alam-ash-saki-ghosh-qaoa-compilation} design a compilation tool specifically for QAOA circuits.
They first assign gates to layers, but due to the commutation relations, most of the dependencies can be ignored, so they apply a two-level search: at the higher level, the order of the layers is decided, followed by a lower level breadth-first search to find the SWAP gates.

\textit{Benchmarks:} To evaluate these approaches to LSQC, benchmarks such as reversible functions in RevLib~\cite{ismvl08-wille-grobe-teuber-dueck-dreschsler-revlib}, gate optimization results of useful logic functions in Amy et al.~\cite{tcad13-amy-maslov-mosca-roeteller-meetinthemiddle} and Nam et al.~\cite{npjqi18-nam-ross-su-childs-maslov-optimization}, or certain circuits with known optimal, QUEKO~\cite{tc20-tan-cong-optimality}, can be used.
The QUEKO evaluations show $\text{t}|\text{ket}\rangle$~\cite{arxiv2003-sivarajah-dilkes-cowtan-simmons-edgington-duncan-tket} to be a leading industry LSQC tool, but still with large optimality gaps.
This serves as a strong motivation for more research into LSQC tools.

\section{Approach}

In this section, we discuss preprocessing, the objectives, the variable encoding scheme, and the constraints of our proposed optimal layout synthesizer for quantum computing (OLSQ).
Then, we introduce some variations to the notion of time to make the synthesizer transition-based (TB-OLSQ), which greatly increases efficiency with little or no performance degradation.
Lastly, we consider commutation to improve TB-OLSQ for QAOA circuits.

\subsection{Preprocessing}
From the input program, we derive a \textit{collision list} $C$: if two gates $g_l$ and $g_{l'}$, $l<l'$, act on a same logical qubit, then we append $(g_l,g_{l'})$ to $C$.
By default, we use the collision list as the \textit{dependency list} $D$.
Users are also welcome to input their own $D$ based on knowledge of the program.
With the dependency list, we can also derive the longest dependency chain with $O(L^2)$ time.
This serves as a lower bound of depth to the LSQC result because the dependency chain can only be lengthened by the SWAP gates and cannot be shortened in any case.
We will need a time coordinate upper bound $T$ in the formulation.
In the hope of a depth-optimal result, we use the longest dependency chain length as $T$ in the beginning.

We also need to extract some features of the coupling graph.
We compute an \textit{overlapping edge pair set} $O$: $\forall e, e'\in E$, $e'\neq e$, if $e$ and $e'$ share some node, append the pair $(e, e')$ to $O$.
We also compute an edge set $E_p$ for each node $p$: $\forall e\in E$, if $e=(\cdot,p)$ or $e=(p,\cdot)$, we append $e$ to $E_p$.
It is straightforward that $E_p \subset E$ and $\cup_{p\in P}E_p=E$.

\subsection{Encoding Variables}

\begin{itemize}
    \item Mapping $\pi_q^t$: at time $t$, logical qubit $q$ is mapped to the physical qubit $\pi_q^t$, $\pi_q^t\in P$.
    \item Time coordinates $t_l$: gate $g_l$ is being executed at time $t_l$, $0\le t_l \le T-1$.
    \item Space coordinates $x_l$: if $g_l\in G_1$, then logical qubit $g_l.q$ is mapped to physical qubit $x_l$, $x_l\in P$; if $g_l\in G_2$, then the two physical qubits, to which $g_l.q$ and $g_l.q'$ are mapped, are connected by edge $x_l$, $x_l\in E$.
    \item Use of SWAP gate $\sigma_k^t$: if $\sigma_k^t=1$, then there is a SWAP gate on edge $e_k$ and the last time slot it takes is $t$ (as SWAP gates may take multiple time slots); otherwise, $\sigma_k^t=0$.
\end{itemize}

\subsection{Constraints}
Note that we differentiate variable assignment $=$ and comparison $==$.
The latter returns true if and only if the equality holds.
There is an additional parameter $S$ in our model which stands for the number of time slots a SWAP gate requires.
$S$ can be set according to different architectures
We set $S=3$ as default, as seen in Figure~\ref{fig:adder_ibmqx2}.

\subsubsection{Injective Mapping} Different logical qubits should be mapped to different physical qubits at any specific time
\begin{equation}
    \pi_{q}^t \neq \pi_{q'}^t \quad \text{for}\ 0 \leq t\leq T-1,\ q,q'\in Q\ \text{and}\ q\neq q' \label{eqn:injective}
\end{equation}

\subsubsection{Avoiding Collisions and Respecting Dependencies}
\begin{equation}
    t_{l} < t_{l'} \quad\text{for}\ (g_l,g_{l'})\in D \label{eqn:respecting}
\end{equation}

\subsubsection{Consistency between Mapping and Space Coordinates} There are two ways we can derive where a gate $g_l$ is at physically: 1) directly through its space coordinate $x_l$; 2) indirectly from the mapping of the logical qubit(s) it acts on at its time coordinate, i.e., $\pi_{g_l.q}^{t_l}$ for single-qubit gates; these two should be consistent.
\begin{equation}
    \left(t_{l}==t\right) \Rightarrow\left(\pi_{g_l.q}^{t}==x_{l}\right) \quad \text{for}\ 0 \leq t \leq T-1, g_l \in G_{1} \label{eqn:consistency_single}
\end{equation}
\begin{equation}
\begin{split}
    \left[\left(t_{l}==t\right)\right.&\wedge\left.\left(x_{l}==e\right)\right] \Rightarrow\\
    &\left\{\left[\left(\pi_{g_l.q}^{t}==e.p\right)\wedge\left(\pi_{g_l.q'}^{t}==e.p'\right)\right]\right.\vee \\
    &\left.\left[\left(\pi_{g_l.q}^{t}==e.p'\right)\wedge\left(\pi_{ g_l.q'}^{t}==e.p\right)\right]\right\} \\
    &\text{for}\ 0 \leq t \leq T-1,\ g_l \in G_{2}, \ e\in E \label{eqn:consistency_two}
\end{split}
\end{equation}

\subsubsection{Proper SWAP Insertion}
Since a SWAP gate takes $S$ time slots, before time $S-1$, no SWAP gates can finish:
\begin{equation}
    \sigma_{k}^t=0 \quad\text{for}\ 0\leq t \leq S-2,\ 0\leq k \leq K-1 \label{eqn:swap_initial}
\end{equation}
A SWAP gate cannot overlap with other SWAP gates on the same edge:
\begin{equation}
\begin{split}
    \left(\sigma_{k}^{t}==1\right) &\Rightarrow\left(\sigma_{k}^{t'}==0\right) \quad \text{for}\ S-1 \leq t \leq T-1,\\
    & t-S+1 \leq t' \leq t-1,\ 0 \leq k \leq K-1\label{eqn:swap_self}
\end{split}
\end{equation}
If two edges overlap in space, the SWAP gates on them cannot overlap in time:
\begin{equation}
\begin{split}
    \left(\sigma_{k}^{t}==1\right) &\Rightarrow\left(\sigma_{k'}^{t'}==0\right) \quad \text{for}\ S-1 \leq t \leq T-1,\\
    & t-S+1 \leq t' \leq t,\ (e_k, e_{k'})\in O \label{eqn:swap_overlapping_edge}
\end{split}
\end{equation} 
A SWAP gate should not overlap with any input single-qubit gates at any time:
\begin{equation}
\begin{split}
    &\left\{\left(t_{l}==t'\right)\wedge\left[\left(x_{l}==e_{k}.p\right) \vee\left(x_{l}==e_{k}.p'\right)\right]\right\} \Rightarrow \left(\sigma_{k}^{t}==0\right)\\
    &\text{for}\ S-1 \leq t \leq T-1,\ t-S+1 \leq t' \leq t,\ 0 \leq k \leq K-1,\ g_l \in G_{1} \label{eqn:swap_input_single}
\end{split}
\end{equation}
A SWAP gate on $e_k$ should not overlap with any input two-qubit gates on the same edge or the edges that overlap at any time:
\begin{equation}
\begin{split}
    \left[\left(t_{l}==t^{\prime}\right) \wedge\left(x_{l}==e_{k'}\right)\right] \Rightarrow\left(\sigma_{k}^{t}==0\right) \quad &\text{for}\ g_l \in G_{2},\\
    S-1 \leq t \leq T,\ t-S+1 \leq t' \leq t,\ (e_k, e_{k'})\in O&\ \text{or}\ k'=k \label{eqn:swap_input_two}
\end{split}
\end{equation}

\subsubsection{Mapping Transformations by SWAP Gates}
Mapping at the next time slot is the same with the current one if there are no SWAP gates finished on all the edges in the edge set $E_p$:
\begin{equation}
    \begin{split}
    &\left[\left(\pi_{q}^{t}==p\right) \wedge\left(\bigwedge_{e_k \in E_{p}} \sigma_{k}^{t}==0\right)\right] \Rightarrow\left(\pi_{q}^{t+1}==p\right) \\
    &\text{for}\ 0 \leq t \leq T-2,\ p\in P,\ q\in Q \label{eqn:transform_no}
    \end{split}
\end{equation}
If there is a SWAP gate finished at $t$, there can only be one. 
(Otherwise, the two SWAP gates are on two edges that overlap.
This case would be ruled out by Equation~\ref{eqn:swap_overlapping_edge}.)
The mapping at $t+1$ is then transformed by the SWAP gate:
\begin{equation}
    \begin{split}
    &\left[\left(\pi_{q}^{t}==e_{k}.p\right) \wedge\left(\sigma_{k}^{t}==1\right)\right] \Rightarrow\left(\pi_{q}^{t+1}==e_{k}.p'\right) \\
    &\left[\left(\pi_{q}^{t}==e_{k}.p'\right) \wedge\left(\sigma_{k}^{t}==1\right)\right] \Rightarrow\left(\pi_{q}^{t+1}==e_{k}.p\right) \\
    &\text{for}\ 0 \leq t \leq T-2,\ 0 \leq k \leq K-1,\ q\in Q
    \label{eqn:transform_yes}
    \end{split}
\end{equation}

\subsection{Objectives}
With the set of variables defined above, it is easy to construct the common objectives. 
In fact, as long as a quantity can be defined from the above variables, it can be the objective.
\textbf{1)} Depth: $d\coloneqq\max_{0,1,...,L-1}t_l$.
We do not need to consider the time coordinates of the inserted SWAP gates because, if a SWAP gate has even larger time coordinate than $d$ defined above, it finishes after all the input gates, thus has no effects on the program and should be ignored.
\textbf{2)} SWAP cost: $c\coloneqq\sum_{k=0}^{K-1}\sum_{t=0}^{T-1} \sigma_{k}^t$.
\textbf{3)} ($\log$-) Fidelity:
\begin{equation}
\begin{split}
    f\coloneqq&\sum_{p \in P} \log f_0(p) \left[\sum_{q\in Q}(\pi_q^{T-1}==p)\right] +\sum_{p\in P} \log f_1(p) \left[\sum_{g_l\in G_1}(x_l==p)\right]  \\
    &+ \sum_{e\in E} \log f_2(e) \left[ \sum_{g_l\in G_2}(x_l==e) \right] + \sum_{k=0}^{K-1} \log f_{SWAP}(e_k) \left[\sum_{t=0}^{T-1}\sigma_{k}^t\right] \label{eqn:fidelity}
\end{split}
\end{equation}
$f_0$, $f_1$, and $f_2$ above are given as input to the LSQC problem.
$f_{SWAP}(e)$ is the fidelity of a SWAP gate on edge $e$, which should be computed from the provided single-qubit and two-qubit gate fidelity, depending on how SWAP gates are implemented on the specific architecture.
In our case, a SWAP gate consists of three $CX$ gates, so $\log f_{SWAP} = 3\log f_2$.
To use addition rather than multiplication in the objective, we take the $\log$ of $f_0$, $f_1$, and $f_2$.
To be compatible with other data and variables, which are all integer, we scale up every $\log$ fidelity value by 1000 and round it to the nearest integer.

\subsection{Complexity Analysis}
In our formulation, there are in total $MT+2L+KT+2N+2K$ variables.
For regular planar graphs, which most coupling graphs are, the number of edges is usually asymptotically linear to the number of nodes.
For example, in a grid, each edge connects two nodes and each node spans out four edges, so $K/N\approx 2$.
Therefore, the total number of variables in our formulation is $O(NT)$ where $N$ is the physical qubit count and $T$ is the time coordinate upper bound.
The total search space is then exponential to $N$ and $T$.
This is expected from the NP-completeness of the problem~\cite{cgo18-siraichi-santos-collange-pereira-allocation, tc20-tan-cong-optimality}.
However, as shown in Table~\ref{tab:complexity}, this formulation still has exponentially fewer variables compared to Wille et al.~\cite{dac19-wille-burgholzer-zulehner-minimal} because they have a variable for each permutation of qubits at each time.
OLSQ also has polynomially fewer variables than Bhattacharjee et al.~\cite{iccad19-bhattacharjee-saki-alam-chattopadhyay-ghosh-muqut}, because they also require variables encoding the mapping from \textit{pairs of} logical qubits to pairs of physical qubits at each time.
Also, it is straightforward from Table~\ref{tab:complexity} that our formulation reduces the number of constraints significantly.

\begin{table}[tb]
    \caption{Complexity of OLSQ and Related Works}
    \vspace{-1\baselineskip}
    \label{tab:complexity}
    \begin{minipage}{\linewidth}
        \begin{center}
           \begin{tabular}{llll}
            \toprule
             &Solver &Variables &Constraints \\
            \midrule
            \cite{dac19-wille-burgholzer-zulehner-minimal} &SMT &$O(L_2\cdot N!)$ &$O(L_2\cdot M\cdot N!)$\\
            \cite{iccad19-bhattacharjee-saki-alam-chattopadhyay-ghosh-muqut} &ILP &$O(T\cdot (N^4 + I))$ & $O(T\cdot (N^4 + I^2 + N\cdot L))$\\
            OLSQ &SMT &$O(T\cdot N + L)$ &$O(T\cdot N\cdot L)$\\
            TB-OLSQ &SMT &$O(B\cdot N + L)$ &$O(B\cdot N\cdot L)$\\
            \bottomrule
            \end{tabular}
        \end{center}
        \footnotesize
        \emph{Note:} $L_2$ is the two-qubit gate count. $N$ is the physical qubit count. $T$ is the time coordinate upper bound. $I$ is the number of levels of gates. $L$ is the total gate count. $B$ is the number of gate blocks. 
    \end{minipage}
\end{table}

\subsection{Optimality of OLSQ}
After we passed the variables, one of the objectives, and the constraints to Z3 SMT solver~\cite{book08-de-moura-bjorner-z3}, it would either return a model containing all the variable values that optimizes the given objective, or return `unsatisfiable'.
As mentioned before, we initially set the time coordinate upper bound $T$ to the largest length of dependency chain.
However, it may be the case that on the given architecture, it is impossible to find a solution with this upper bound, e.g. LSQC of the quantum adder, shown in Figure~\ref{fig:adder_diagram}, on IBM QX2 coupling graph, shown in Figure~\ref{fig:coupling_graph_ibmqx2}.
Thus, if the model is unsatisfiable, we geometrically increase $T$ each time by $(1+\epsilon)$x until it is satisfiable.
We set $\epsilon=0.3$ in our experiment.
This means that OLSQ is optimal up to a certain time coordinate upper bound $T$.
For depth optimization, the optimality is guaranteed.
However, for SWAP cost and fidelity optimization, sometimes increasing $T$ even more can lead to better results.
However, this is a very rare case as we shall see in Section~\ref{sec:eva-olsq}, especially when the longest dependency chain in the input quantum program is already of considerable length.

\subsection{TB-OLSQ: Rethinking Time Coordinates}

\begin{figure}[tb]
    \centering
    \includegraphics[width=\columnwidth]{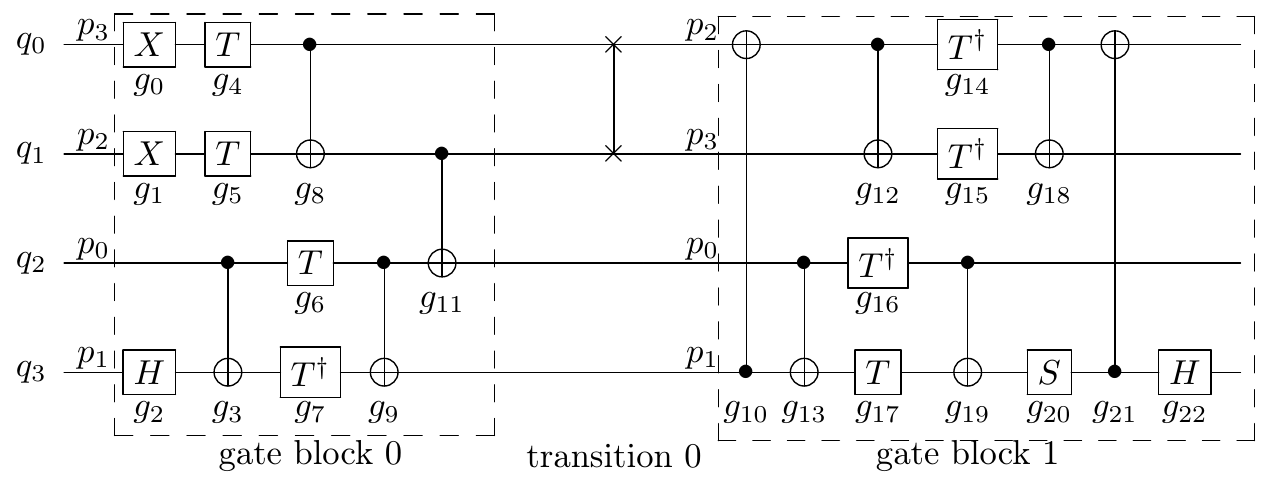}
    \vspace{-1.5\baselineskip}
    \caption{Quantum adder in transition-based model (Two `x's connected by a vertical line segment represent a SWAP gate.)}
    \label{fig:adder_tbOLSQ}
\end{figure}

In the example of quantum adder, the time upper bound is $T=15$.
However, the mapping to physical qubits changed only once at time $8$.
Thus, for any specific logical qubit $q$, the variables $\pi_q^t$ are the same from $t=0$ to $7$ and $t=8$ to $14$.
This is a huge redundancy.
In addition, the total search space for the solver is exponential to $N$ and $T$.
Although a cutting edge quantum processor has only $N=53$~\cite{nature19-google-supremacy}, $T$ as determined by quantum programs can easily grow to be quite large.

These two observations motivate us to improve efficiency of OLSQ by rethinking time coordinates.
Instead of keeping $\pi_q^t$ for all $t$, it turns out that we can keep only these variables between two \textit{transitions} of the mapping.
Formally, a transition is a set of parallel SWAP gates.
In the quantum adder example, there is only one transition and the transition consists of only one SWAP gate on edge $e_3=(p_2,p_3)$.
SWAP gates on overlapping edges can not be in parallel, according to Equation~\ref{eqn:swap_overlapping_edge}, so $e_1$, $e_2$, $e_4$, and $e_5$ cannot be in the same transition with $e_3$.
However, $e_0$ and $e_3$ together is a valid transition.

Now, we consider a new model of execution which separates the input gates and the inserted SWAP gates: executing some input gates, then a few SWAP gates to make transition(s) in mapping, execute some more input gates, and make other transition(s), ...
We can have consecutive transitions without executing any input gates in between.
This model is similar to the one in Wille et al.~\cite{dac19-wille-burgholzer-zulehner-minimal}, but gates later in the input can appear at the front as long as permitted by dependency, which they do not allow.
The quantum adder example in this model is shown in Figure~\ref{fig:adder_tbOLSQ}, where we first execute the gates in gate block $0$, then a SWAP gate to make transition $0$, finally the gates in gate block $1$.
In this transition-based model, there is no notion of precise time.
Instead, all the gates in a gate block are at a same coarse-grain time slot, and share the same mapping.
This way, the number of mapping variables greatly decreases.
In this particular example, there are only $8$ mapping variables compared to the original $60$ mapping variables.

With slight changes of formulation, OLSQ can be made transition-based.
We change the $<$ in Equation~\ref{eqn:respecting} to $\leq$, since now even if two input gates have dependency, they can still be assigned to the same gate block, meaning their coarse-grain time coordinates can still be the same.
We set $S=1$ and remove Equation~\ref{eqn:swap_input_single} and Equation~\ref{eqn:swap_input_two}.
SWAP gates are separated from input gates in the current model, so we do not need to consider overlaps of input gates with them.
$S=1$ because we are using a coarse-grain time model.

The coarse-grain time upper bound $T$ is initially set to $1$, so the solver will search for a solution without any transition.
If the solver returns `unsatisfiable', we will increase $T$ by $1$ each time until it finds a solution that optimizes the given objective.
The value of depth would just be $T-1$, since there are exactly $T-1$ gate blocks in the resulting circuit.
If SWAP cost or fidelity is set as objective, TB-OLSQ will find the optimal solution that has up to $T-1$ transitions.
Just like OLSQ, there may be better solutions if $T$ is increased even more.
After a solution such as Figure~\ref{fig:adder_tbOLSQ} is returned, we can use as-soon-as-possible (ASAP) scheduling to derive all the exact time coordinates of the gates.
After scheduling, the resulting format is exactly the same as that of OLSQ.
We shall show that TB-OLSQ produced optimal or near-optimal solution with orders-of-magnitude speedup compared to OLSQ.

\subsection{QAOA-OLSQ: Removing False Dependencies}

\begin{figure}[tb]
    \begin{minipage}[b]{0.6\columnwidth}
        \includegraphics[width=\linewidth]{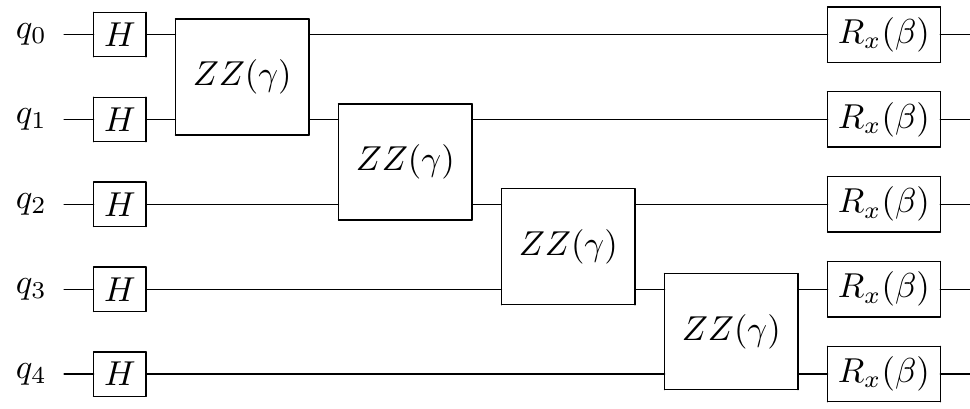}
        \vspace{-1\baselineskip}
        \subcaption{Original QAOA program}
        \label{fig:qaoa}
    \end{minipage}
    \hfill
    \begin{minipage}[b]{0.37\columnwidth}
        \includegraphics[width=\linewidth]{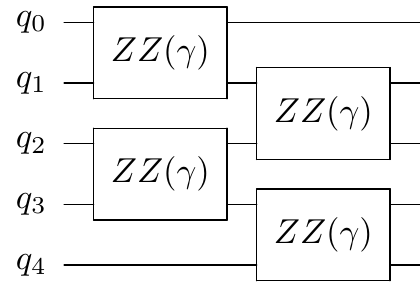}
        \vspace{-1\baselineskip}
        \subcaption{Better phase separation}
        \label{fig:qaoa_OLSQ}
    \end{minipage}
    \vspace{-1\baselineskip}
    \caption{}
\end{figure}

One of the most important concepts in quantum mechanics is commutation.
Only when two operators commute, can we simultaneously `observe' them and receive definite values.
The famous Heisenberg's uncertainty principle states that position and momentum of a particle do not commute, so we cannot simultaneously receive definite values of these two.
In the case of quantum computing, if two gates $g_l$ and $g_{l'}$ ($l<l'$) commute, we can change their relative order without altering the whole program.
This means that even if they consecutively act on the same qubit $q_m$, $t_l$ is not necessarily smaller than $t_{l'}$.
Since commutation is purely logical and has nothing to do with the architecture, one may think that it is solely the job of logic synthesis to experiment with the commutation relations.
Like many other related works, both OLSQ and TB-OLSQ assume that any commutation is performed prior to LSQC and thus will indeed add $(g_l,g_{l'})$ as a dependency, eliminating the possibility of $t_l>t_{l'}$.
However, it turns out that more knowledge of dependencies in LSQC is very beneficial --- especially on the QAOA programs~\cite{arxiv1411-farhi-goldstone-gutmann-qaoa, algorithms19-hadfield-wang-ogorman-rieffel-venturelli-biswas-qaoa}, which shows great promise to solve some approximate discrete optimization problems.

The QAOA program first sets the qubits to the equal superposition state by applying Hadamard $H$ gates on all the qubits.
Then it goes into many iterations.
Each iteration has two stages: phase separation and mixing.
A simple QAOA program with only one iteration is shown in Figure~\ref{fig:qaoa}~\cite{arxiv2004-gokhale-javadi-abhari-earnest-shi-chong-openpulse}.
Phase separation is implemented by a few $ZZ$ gates, which are two-qubit gates with a parameter $\gamma$; mixing is implemented by $R_x$ gates on all the qubits.
The specific functions of these gates are not of concern to Layout synthesis; what matters is on which qubit(s) they act.
Since single-qubit gates are always executable, the mixing stage does not require layout synthesis.
However, phase separation may contain a $ZZ$ gate on any pair of qubits, which means that layout synthesis is required to move the non-adjacent qubits together when a $ZZ$ gate needs to act on them.

If we input the QAOA program, Figure~\ref{fig:qaoa}, and IBM QX2 coupling graph, Figure~\ref{fig:coupling_graph_ibmqx2}, to OLSQ or TB-OLSQ, the best result is probably just an identity mapping and the output looks the same with Figure~\ref{fig:qaoa}.
OLSQ and TB-OLSQ cannot reduce depth because the four $ZZ$ gates all depend on the gate before it, imposing the default dependencies.
The particularity of QAOA is that all the $ZZ$ gates commute, which means even both $ZZ(\gamma)(q_2, q_3)$ and $ZZ(\gamma)(q_3, q_4)$ act on $q_3$, the latter can actually commute `through' the former.
As a result, we have a better phase separation subroutine, as shown in Figure~\ref{fig:qaoa_OLSQ}, which has smaller depth.
Current qubits still have short `lifetimes' and many QAOA applications have large numbers of iterations, so reducing depth is crucial.

We can improve TB-OLSQ for the phase separation stage of QAOA with the knowledge of commutation.
Previously, we treat every collision as a dependency.
However, in the phase separation stage, all the $ZZ$ gates are commutable, so none of the collisions are real dependencies.
Thus, we simply remove the constraints in Equation~\ref{eqn:respecting} in TB-OLSQ.
Up to this point, the result would be blocks of original $ZZ$ gates with the fewest transitions possible to make all the qubit pairs adjacent required by these $ZZ$ gates.
Inside the $ZZ$ gate blocks, there may be further opportunities to reduce depth with the help of commutation.
Therefore, we input this result to OLSQ with depth as objective and, again, remove the constraints in Equation~\ref{eqn:respecting}.
Since the gates are already mapped to valid edges on the coupling graph, OLSQ does not need to insert any new SWAP gates.
Thus, we disable all the $\sigma_k^t$ variables in OLSQ for speedup.
In the end, we derive the spacetime coordinates of all the $ZZ$ gates and the SWAP gates inserted.
The depth of this result is highly optimized by the two passes of TB-OLSQ and OLSQ.

\section{Evaluation}

The evaluations were run on an Ubuntu 16.04 server with two Intel Xeon E5-2699v3 CPUs and 128GB memory.
Wille et al.~\cite{dac19-wille-burgholzer-zulehner-minimal}\footnote{\url{https://github.com/iic-jku/minimal_ibm_qx_mapping}} and TriQ~\cite{isca19-murali-linke-martonosi-abhari-nguyen-triq}\footnote{\url{https://github.com/prakashmurali/TriQ}} were built with Cmake 3.13.4 and GNU Make 4.1.
The version of Python was 3.8.2.
The versions of Python packages used were Cirq 0.8.0, Pytket 0.5.4, Qiskit 0.18.0, and Z3-Solver 4.8.7.0.
We linked the Z3 library contained in Z3-Solver package to Wille et al. and TriQ when building.

We selected a comprehensive set of benchmarks from various sources including \cite{tcad13-amy-maslov-mosca-roeteller-meetinthemiddle, npjqi18-nam-ross-su-childs-maslov-optimization, arxiv2004-gokhale-javadi-abhari-earnest-shi-chong-openpulse,tc20-tan-cong-optimality}.
We used the fidelity profile of IBM QX2, Figure~\ref{fig:coupling_graph_ibmqx2}, and IBM Melbourne~\ref{fig:coupling_graph_melbourne}, from~\cite{isca19-murali-linke-martonosi-abhari-nguyen-triq}.
To evaluate fidelity, we inputted the result from different synthesizers to Qiskit, decomposed, and calculated the product of all gate fidelity.
We made all the benchmarks and detailed results open-source.\footnote{\url{https://github.com/UCLA-VAST/OLSQ}}.

\begin{figure}[tb]
    \begin{tikzpicture}    
        \begin{groupplot}[
            group style={
            group name=asymptotics,
            group size=2 by 1,
            xticklabels at=edge bottom,
            horizontal sep=2em
            },
            ymode=log,
            ymin=0.1, ymax=4e4,
            height=15em,
            ylabel=$t$/s
            ]
    
        \nextgroupplot[
            xmin=4.5,xmax=8.5,
            xtick={5, 6, 8},
            axis x line=bottom,
            axis y line=middle,
            width=16em,
            xlabel={Number of Physical Qubits},
            title={Benchmark 4mod5-v1\_22},
            legend cell align={left},
            legend style={font=\footnotesize, 
                          at={(axis cs:5.1, 1e4)}, anchor=north west
                         }
            ]
        
        \addplot[
        color=blue,
        only marks,
        mark=triangle*,
        ultra thick
        ] 
            coordinates {
            (5, 0.806)
            (6, 1.826)
            (8, 946.656)};
        
        \addplot[
        color=red,
        only marks, 
        mark=square*, 
        ultra thick
        ] 
            coordinates {
            (5, 44.53523063659668)
            (6, 10.782002925872803)
            (8, 14.661139249801636)
            };
        
        \addplot [
        domain=4.7:8.3, 
        samples=100,
        dashed,
        thick
        ]
        {exp(2.46609*x - 13.2054)};
    
        \nextgroupplot[
        xmin=4.5,xmax=8.5,
        xtick={5, 6, 8},
        axis x line=bottom,
        axis y line=middle,
        width=16em,
        xlabel={Number of Physical Qubits},
        title={Benchmark adder},
        legend cell align={left},
        legend style={font=\footnotesize, 
                      at={(axis cs:8.5, 0.15)}, anchor=south east
                     }
        ]
        
        \addplot[
        color=blue,
        only marks,
        mark=triangle*,
        ultra thick
        ] 
            coordinates {
            (5, 0.509)
            (6, 8.318)
            (8, 10315.601)};
        
        \addplot[
        color=red, 
        only marks, 
        mark=square*, 
        ultra thick
        ] 
            coordinates {
            (5, 22.54019284248352)
            (6, 413.7794051170349)
            (8, 714.0979297161102)};
        
        \addplot [
        domain=4.7:8.2, 
        samples=100,
        dashed,
        thick
        ]
        {exp(-17.6053 + 3.34213*x)};
        
        \legend{Wille et al., OLSQ-SWAP}
        \end{groupplot}
    \end{tikzpicture}
    \captionsetup{singlelinecheck=off,font=footnotesize}
    \vspace{-2\baselineskip}
    \caption*{\textit{Note:} The devices with 5, 6, 8, and 16 qubits are IBM QX2, a 2 by 3 grid, a 2 by 4 grid, and Rigetti Aspen-4. Dashed line is an exponential fit of Wille et al.'s results.}
    
    \captionsetup{singlelinecheck=on,font=normalsize}
    \vspace{-0.5\baselineskip}
    \caption{Runtime Scaling of Wille et al., and OLSQ-SWAP}
    \label{fig:asymptotics}

\end{figure}

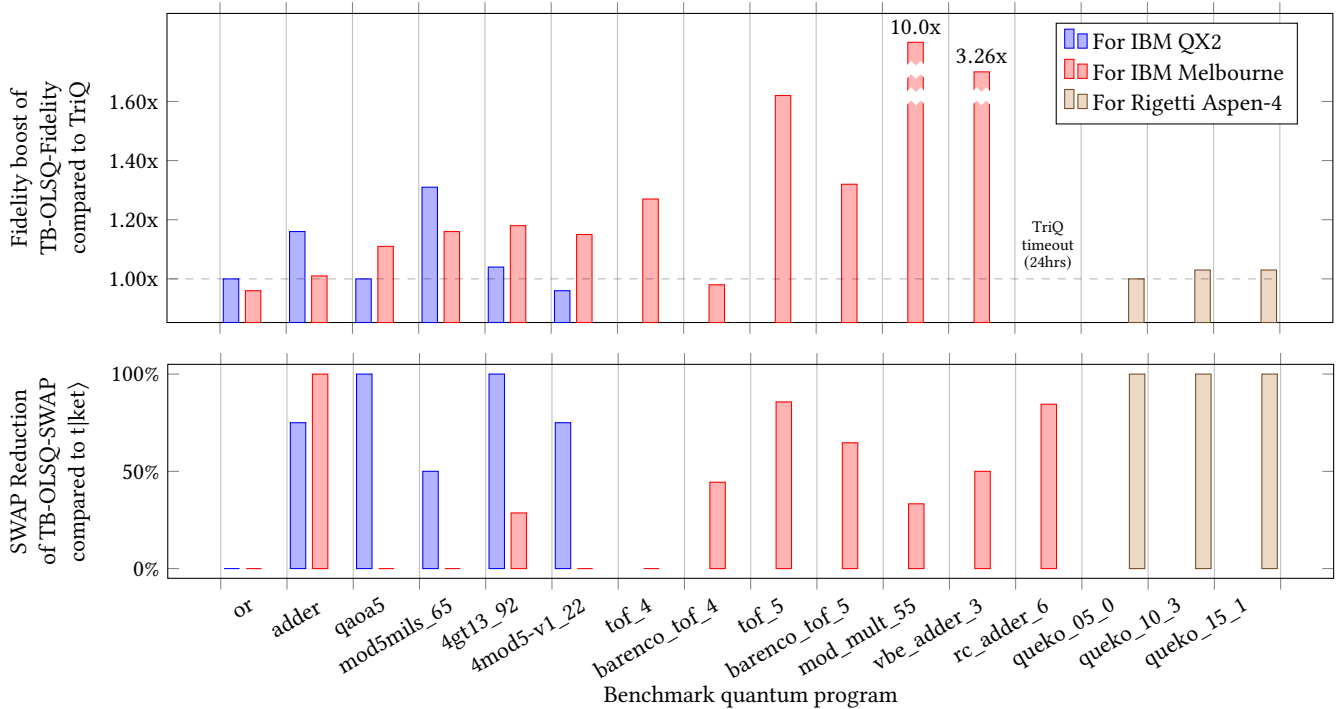
\begin{figure*}[tb]
    \begin{tikzpicture}
            \begin{axis}[
            	x tick label style={/pgf/number format/1000 sep=},
            	ybar interval=0.7,
            	enlargelimits=0.05,
            	ymin=0.9,ymax=1.85,
                     ylabel style={align=center, yshift=1em},
                    ylabel={Fidelity boost of \\ TB-OLSQ-Fidelity \\ compared to TriQ},
                    xmin=1, xmax=17,
                    xtick={1, 2, 3, 4, 5, 6, 7, 8, 9, 10, 11, 12, 13, 14, 15, 16, 17},
                    xticklabels=\empty,
                    width=54em,
                    height=18em,
                    legend pos=north east,
                    legend cell align={left},
                    ytick={1, 1.2, 1.4, 1.6},
                    yticklabels={1.00x, 1.20x, 1.40x, 1.60x}
            ]
            
            \addplot 
            coordinates {
                (1, 1)
                (2, 1.16)
                (3, 1)
                (4, 1.31)
                (5, 1.04)
                (6, 0.96)
                (7,0)
            };
            
            \addplot 
            coordinates {
                (1, 0.96)
                (2, 1.01)
                (3, 1.11)
                (4, 1.16)
                (5, 1.18)
                (6, 1.15)
                (7, 1.27)
                (8, 0.98)
                (9, 1.62)
                (10, 1.32)
                (11, 1.8)
                (12, 1.7)
                (13,0)
            };
            
            \addplot
            coordinates {
                (14, 1)
                (15, 1.03)
                (16, 1.03)
                (17,0)
            };
            
            \legend{For IBM QX2, For IBM Melbourne, For Rigetti Aspen-4}

            \node [above] at (axis cs:  11.5, 1.8) {$\mathrm{10.0x}$};
            \node [above] at (axis cs:  12.5, 1.7) {$\mathrm{3.26x}$};
            \node [above] at (axis cs:  13.5, 1.12) {\small $\scriptstyle\mathrm{TriQ}$};
            \node [above] at (axis cs:  13.5, 1.07) {\small $\scriptstyle\mathrm{timeout}$};
            \node [above] at (axis cs:  13.5, 1) {\small $\scriptstyle\mathrm{(24hrs)}$};
            
            \draw[dashed, opacity=0.3] (axis cs: 0,1) -- (axis cs:17,1);
            \end{axis}
            
            \draw[color=white, fill=white] (10.7, 2.9) --++(0.2em, 0.2em) --++ (0.2em, -0.2em) --++(0.2em, 0.2em) --++ (0.2em, -0.2em) --++ (0, 0.5em) --++ (-0.2em, 0.2em) --++ (-0.2em, -0.2em) --++ (-0.2em, 0.2em) --++ (-0.2em, -0.2em) -- cycle;
            
            \draw[color=white, fill=white] (9.83, 2.9) --++(0.2em, 0.2em) --++ (0.2em, -0.2em) --++(0.2em, 0.2em) --++ (0.2em, -0.2em) --++ (0, 0.5em) --++ (-0.2em, 0.2em) --++ (-0.2em, -0.2em) --++ (-0.2em, 0.2em) --++ (-0.2em, -0.2em) -- cycle;
            
            \draw[color=white, fill=white] (9.83, 3.25) --++(0.2em, 0.2em) --++ (0.2em, -0.2em) --++(0.2em, 0.2em) --++ (0.2em, -0.2em) --++ (0, 0.5em) --++ (-0.2em, 0.2em) --++ (-0.2em, -0.2em) --++ (-0.2em, 0.2em) --++ (-0.2em, -0.2em) -- cycle;
        
        \end{tikzpicture}

    \begin{tikzpicture}
        \begin{axis}[
        	x tick label style={/pgf/number format/1000 sep=},
        	enlargelimits=0.05,
        	ybar interval=0.7,
        	ymin=0,ymax=1,
                 ylabel style={align=center, yshift=1.065em},
                ylabel={SWAP Reduction \\ of TB-OLSQ-SWAP \\ compared to $\text{t}|\text{ket}\rangle$},
                xlabel={Benchmark quantum program},
                xlabel style={yshift=-2.5em},
                xmin=1, xmax=17,
                xtick={1, 2, 3, 4, 5, 6, 7, 8, 9, 10, 11, 12, 13, 14, 15, 16,17},
                xticklabels={or, adder, qaoa5, mod5mils\_65, 4gt13\_92, 4mod5-v1\_22, tof\_4, barenco\_tof\_4, tof\_5, barenco\_tof\_5, mod\_mult\_55, vbe\_adder\_3, rc\_adder\_6, queko\_05\_0, queko\_10\_3, queko\_15\_1,na},
                xticklabel style={rotate=30, anchor=east, yshift=-0.5em},
                width=54em,
                height=14em,
                legend pos=outer north east,
                legend cell align={left},
                ytick={0, 0.5, 1},
                yticklabels={0\%, 50\%, 100\%}
        ]

        \addplot
        coordinates {
            (1, 0)
            (2, 0.75)
            (3, 1)
            (4, 0.5)
            (5, 1)
            (6, 0.75)
            (7,0)
        };
        
        \addplot 
        coordinates {
            (1, 0)
            (2, 1)
            (3, 0)
            (4, 0)
            (5, 0.286)
            (6, 0)
            (7, 0)
            (8, 0.444)
            (9, 0.857)
            (10, 0.647)
            (11, 0.333)
            (12, 0.500)
            (13, 0.845)
            (14,0)
        };
        
        \addplot
        coordinates {
            (14, 1)
            (15, 1)
            (16, 1)
            (17,0)
        };
        
        \end{axis}
    \end{tikzpicture}
    \vspace{-2\baselineskip}
    \caption{Evaluation of TB-OLSQ}
    \label{fig:eva-tb-olsq}
\end{figure*}

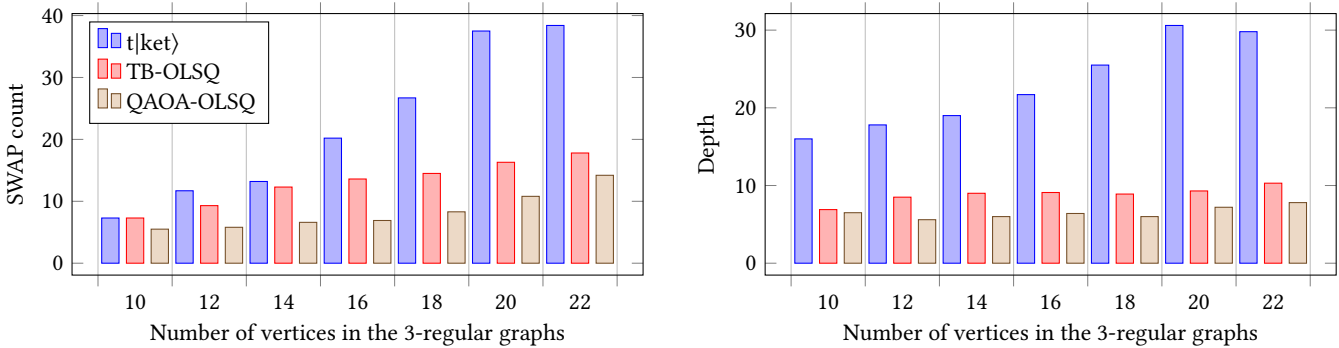
\begin{figure*}[tb]
    \begin{tikzpicture}
            \begin{axis}[
            	x tick label style={/pgf/number format/1000 sep=},
            	ylabel=SWAP count,
            	enlargelimits=0.05,
            	ybar interval=0.7,
            	legend pos=north west,
            	xlabel=Number of vertices in the 3-regular graphs,
            	ylabel style={yshift=-1.5em},
            	width=29em,
            	height=16em,
            	legend cell align={left}
            ]
            \addplot 
            	coordinates {
            	(10, 7.3)
            	(12, 11.7)
            	(14, 13.2)
            	(16, 20.2)
            	(18, 26.7)
            	(20, 37.5)
            	(22, 38.4)
            	(24,0)
            	};
            	
            \addplot 
            	coordinates {
            	(10, 7.3)
            	(12, 9.3)
            	(14, 12.3)
            	(16, 13.6)
            	(18, 14.5)
            	(20, 16.3)
            	(22, 17.8)
            	(24,0)
            	};
            	
            \addplot 
            	coordinates {
            	(10, 5.5)
            	(12, 5.8)
            	(14, 6.6)
            	(16, 6.9)
            	(18, 8.3)
            	(20, 10.8)
            	(22, 14.2)
            	(24,0)
            	};
            	
            \legend{$\text{t}|\text{ket}\rangle$, TB-OLSQ, QAOA-OLSQ}
            \end{axis}
        \end{tikzpicture}
    \hfill
        \begin{tikzpicture}
            \begin{axis}[
            	x tick label style={/pgf/number format/1000 sep=},
            	ylabel=Depth,
            	enlargelimits=0.05,
            	ybar interval=0.7,
            	legend pos=north west,
            	xlabel=Number of vertices in the 3-regular graphs,
            	ylabel style={yshift=-1.5em},
            	width=29em,
            	height=16em,
                legend cell align={left}
            ]
            \addplot 
            	coordinates {
            	(10, 16)
            	(12, 17.8)
            	(14, 19)
            	(16, 21.7)
            	(18, 25.5)
            	(20, 30.6)
            	(22, 29.8)
            	(24,0)
            	};
            	
            \addplot 
            	coordinates {
            	(10, 6.9)
            	(12, 8.5)
            	(14, 9.0)
            	(16, 9.1)
            	(18, 8.9)
            	(20, 9.3)
            	(22, 10.3)
            	(24,0)
            	};
            	
            \addplot 
            	coordinates {
            	(10, 6.5)
            	(12, 5.6)
            	(14, 6)
            	(16, 6.4)
            	(18, 6)
            	(20, 7.2)
            	(22, 7.8)
            	(24,0)
            	};
            	
            \end{axis}
        \end{tikzpicture}
        
    \vspace{-1\baselineskip}
    \caption{Evaluation of QAOA-OLSQ}
    \vspace{-1\baselineskip}
    \label{fig:qaoa_result}
\end{figure*}

\subsection{OLSQ versus Previous Optimal Approaches}
\label{sec:eva-olsq}

The leading exact approach by Wille et al.~\cite{dac19-wille-burgholzer-zulehner-minimal} is open-source.
The coupling graph in their original paper is directed, which means that $CX$ gate can only execute in one direction.
However, the edges in our formulation is bi-directional.
For fair comparison, we input each bi-directional edge as two uni-directional edges to their software.
We observe that, compared to the cases where directed coupling graphs are used, their software runs significantly faster.
Thus, the runtime data we collect, for comparison with OLSQ, are smaller than those appeared in the original reference~\cite{dac19-wille-burgholzer-zulehner-minimal}.
Since Wille et al. aims to minimize SWAP cost, it is most appropriate to compare it against OLSQ with SWAP cost as objective, denoted as OLSQ-SWAP below.
We find that, on all instances of benchmark program and architecture, OLSQ-SWAP matches the performance of Wille et al. and sometimes is even better.

Wille et al. have a set of variables denoting whether each qubit permutation is performed between two two-qubit gates.
This means that the number of variables in their formulation is proportional to $N!$, where $N$ is the physical qubit count.
This complexity explosion can be observed from the two examples in Figure~\ref{fig:asymptotics}, where the runtime of Wille et al. is nearly exponential.
On the other hand, the runtime of OLSQ-SWAP does not show this exponential growth.
Wille et al. also rely on pre-processing to derive a function from each qubit permutation to its SWAP cost.
Suppose each function value takes 4B memory, then, even for device with qubit count $N=12$, the memory required is 2TB.
As examples, we evaluate three LSQC instances of QUEKO benchmarks and the 16-qubit architecture Aspen-4, Figure~\ref{fig:coupling_graph_aspen4}.
All these evaluations of Wille et al. are aborted because our server runs out of the memory limit we set (32GB).
In contrast, OLSQ-SWAP reach the optimal SWAP cost (0) within a relatively short period of time.

\subsection{TB-OLSQ versus Heuristic Approaches}

One of the leading industry works, $\text{t}|\text{ket}\rangle$~\cite{arxiv2003-sivarajah-dilkes-cowtan-simmons-edgington-duncan-tket}, mainly focuses on optimizing cost.
We compare it with TB-OLSQ with SWAP cost set as objective, denoted as TB-OLSQ-SWAP below.
One of the leading academic works, TriQ~\cite{isca19-murali-linke-martonosi-abhari-nguyen-triq}, mainly focuses on optimizing fidelity, so we compare it to TB-OLSQ with fidelity set as objective, denoted as TB-OLSQ-Fidelity below.
The results are given in Figure~\ref{fig:eva-tb-olsq}.

The evaluations denoted as blue dots are the same LSQC instances given to the optimal approaches, so that we can compare heuristic and exact approaches.
We find that TB-OLSQ-SWAP has no observable degradation on $CX$ cost compared to OLSQ-SWAP, neither does TB-OLSQ-Fidelity compared to OLSQ-Fidelity on fidelity.
This means that our transition-based approach is almost exact; it also hugely increases efficiency, e.g. solving queko\_15\_1 takes the former 9E4 seconds while taking the latter 30 seconds.
Compared to $\text{t}|\text{ket}\rangle$, TB-OLSQ-SWAP often reduces the $CX$ cost by large margins, 69.2\% in geometric mean.
Compared to TriQ, TB-OLSQ-Fidelity often increases the fidelity, even up to 10.0x on some larger programs.
In some cases, TriQ has slightly better fidelity results by leveraging larger depths.
This is expected since TB-OLSQ-Fidelity optimizes fidelity with the fewest transitions possible, but the number of transitions has no direct link to fidelity, so it is possible to find solutions with higher fidelity and more transitions.
However, compared to TriQ, the fidelity loss of OLSQ in these cases remain less than 5\%.

\subsection{QAOA-OLSQ}
\label{sec:qaoa-olsq}

Arute et al.~\cite{arxiv2004-google-qaoa} is considered a leading work on QAOA implementation, where $\text{t}|\text{ket}\rangle$ is used to solve LSQC problems of QAOA programs for 3-regular graphs.
We conduct evaluation with the same settings in~\cite{arxiv2004-google-qaoa}: the coupling graph is part of Google Sycamore with 23 physical qubits, as shown in Figure~\ref{fig:coupling_graph_sycamore23}.
We generate random 3-regular graphs with node count $M$ from 10 to 22.
(Graph theory shows that the product of $M$ and the vertex degree must be even, so $M$ must be even.)
Then, for each edge $(i,j)$ in a 3-regular graph, we append a corresponding gate $ZZ(q_i,q_j)$ to the phase separation stage.
The phase separation is given to $\text{t}|\text{ket}\rangle$, TB-OLSQ, and QAOA-OLSQ for LSQC and their results are fed to Cirq for statistics on SWAP cost and depth.
We consider all $ZZ$ gates and SWAP gates to have unity depth.
As shown in Figure~\ref{fig:qaoa_result}, even without considering the commutation relations, TB-OLSQ reduces depth by 59.5\% in geometric mean and reduces SWAP cost by 29.4\% in geometric mean.
QAOA-OLSQ further reduces both depth and SWAP cost, by 70.2\% and 53.8\% in geometric mean compared to $\text{t}|\text{ket}\rangle$.

\section{Conclusion and Future Work}
In this paper, we formulate layout synthesis for quantum computing as optimization problems.
We present two synthesizers: an exact layout synthesizer (OLSQ) and an approximate, transition-based synthesizer (TB-OLSQ).
In addition, we improve TB-OLSQ for QAOA programs (QAOA-OLSQ) by considering commutation.
Compared to previous exact approaches, OLSQ is the first that features guaranteed optimality and efficiency both in time and space for general quantum processors.
TB-OLSQ shows no visible degradation than OLSQ and is able to outperform leading academic and industry works on multiple metrics.
QAOA-OLSQ can reduce both the cost and the depth in a real QAOA experiment setting, which means it would be beneficial for realistic applications for near-term quantum computers.
The basic formulation we produce is very versatile in that it can support different objectives, so in the future it can be used even with more complicated objectives involving, say, correlated errors.
It is also very valuable to improve the synthesizers based on prior knowledge of either the program, e.g., commutation relations, or the coupling graph, e.g., special structures inside the graph.
In an era when quantum computers still are volatile and error-prone, layout synthesizers prove to be vital for the execution of all quantum programs to minimize the depth (equivalent to coherence time improvement) and cost, and maximize the fidelity.

\begin{acks}
This work is partially supported by NEC under the Center for Domain-Specific Computing Industrial Partnership Program.
We would like to thank Iris Cong for valuable comments on the manuscript, and the Google Quantum team (Ryan Babbush, Dave Bacon, Craig Gidney, and Matthew Harrigan) for suggesting the QAOA application.
Jason Cong also appreciates multiple discussions with Eugene Ding on the LSQC problem.
\end{acks}

\clearpage
\bibliographystyle{ACM-Reference-Format}
\bibliography{BochenTan}


\begin{thebibliography}{29}


\ifx \showCODEN    \undefined \def \showCODEN     #1{\unskip}     \fi
\ifx \showDOI      \undefined \def \showDOI       #1{#1}\fi
\ifx \showISBNx    \undefined \def \showISBNx     #1{\unskip}     \fi
\ifx \showISBNxiii \undefined \def \showISBNxiii  #1{\unskip}     \fi
\ifx \showISSN     \undefined \def \showISSN      #1{\unskip}     \fi
\ifx \showLCCN     \undefined \def \showLCCN      #1{\unskip}     \fi
\ifx \shownote     \undefined \def \shownote      #1{#1}          \fi
\ifx \showarticletitle \undefined \def \showarticletitle #1{#1}   \fi
\ifx \showURL      \undefined \def \showURL       {\relax}        \fi
\providecommand\bibfield[2]{#2}
\providecommand\bibinfo[2]{#2}
\providecommand\natexlab[1]{#1}
\providecommand\showeprint[2][]{arXiv:#2}

\bibitem[\protect\citeauthoryear{Alam, Ash-Saki, and Ghosh}{Alam
  et~al\mbox{.}}{2020}]%
        {dac20-alam-ash-saki-ghosh-qaoa-compilation}
\bibfield{author}{\bibinfo{person}{Mahabubul Alam}, \bibinfo{person}{Abdullah
  Ash-Saki}, {and} \bibinfo{person}{Swaroop Ghosh}.}
  \bibinfo{year}{2020}\natexlab{}.
\newblock \showarticletitle{An {Efficient} {Circuit} {Compilation} {Flow} for
  {Quantum} {Approximate} {Optimization} {Algorithm}}. In
  \bibinfo{booktitle}{\emph{2020 57th ACM/IEEE Design Automation Conference
  (DAC)}}. \bibinfo{pages}{6}.
\newblock


\bibitem[\protect\citeauthoryear{{Amy}, {Maslov}, {Mosca}, and
  {Roetteler}}{{Amy} et~al\mbox{.}}{2013}]%
        {tcad13-amy-maslov-mosca-roeteller-meetinthemiddle}
\bibfield{author}{\bibinfo{person}{M. {Amy}}, \bibinfo{person}{D. {Maslov}},
  \bibinfo{person}{M. {Mosca}}, {and} \bibinfo{person}{M. {Roetteler}}.}
  \bibinfo{year}{2013}\natexlab{}.
\newblock \showarticletitle{A Meet-in-the-Middle Algorithm for Fast Synthesis
  of Depth-Optimal Quantum Circuits}.
\newblock \bibinfo{journal}{\emph{IEEE Transactions on Computer-Aided Design of
  Integrated Circuits and Systems}} \bibinfo{volume}{32}, \bibinfo{number}{6}
  (\bibinfo{year}{2013}), \bibinfo{pages}{818--830}.
\newblock


\bibitem[\protect\citeauthoryear{Arute, Arya, Babbush, Bacon, Bardin, Barends,
  Biswas, Boixo, Brandao, Buell, et~al\mbox{.}}{Arute et~al\mbox{.}}{2019}]%
        {nature19-google-supremacy}
\bibfield{author}{\bibinfo{person}{Frank Arute}, \bibinfo{person}{Kunal Arya},
  \bibinfo{person}{Ryan Babbush}, \bibinfo{person}{Dave Bacon},
  \bibinfo{person}{Joseph~C Bardin}, \bibinfo{person}{Rami Barends},
  \bibinfo{person}{Rupak Biswas}, \bibinfo{person}{Sergio Boixo},
  \bibinfo{person}{Fernando~GSL Brandao}, \bibinfo{person}{David~A Buell},
  {et~al\mbox{.}}} \bibinfo{year}{2019}\natexlab{}.
\newblock \showarticletitle{Quantum supremacy using a programmable
  superconducting processor}.
\newblock \bibinfo{journal}{\emph{Nature}} \bibinfo{volume}{574},
  \bibinfo{number}{7779} (\bibinfo{year}{2019}), \bibinfo{pages}{505--510}.
\newblock


\bibitem[\protect\citeauthoryear{Arute, Arya, Babbush, Bacon, Bardin, Barends,
  Boixo, Broughton, Buckley, Buell, Burkett, Bushnell, Chen, Chen, Chiaro,
  Collins, Courtney, Demura, Dunsworth, Farhi, Fowler, Foxen, Gidney, Giustina,
  Graff, Habegger, Harrigan, Ho, Hong, Huang, Ioffe, Isakov, Jeffrey, Jiang,
  Jones, Kafri, Kechedzhi, Kelly, Kim, Klimov, Korotkov, Kostritsa, Landhuis,
  Laptev, Lindmark, Leib, Lucero, Martin, Martinis, McClean, McEwen, Megrant,
  Mi, Mohseni, Mruczkiewicz, Mutus, Naaman, Neeley, Neill, Neukart, Neven, Niu,
  O'Brien, O'Gorman, Ostby, Petukhov, Putterman, Quintana, Roushan, Rubin,
  Sank, Satzinger, Skolik, Smelyanskiy, Strain, Streif, Sung, Szalay,
  Vainsencher, White, Yao, Yeh, Zalcman, and Zhou}{Arute et~al\mbox{.}}{2020}]%
        {arxiv2004-google-qaoa}
\bibfield{author}{\bibinfo{person}{Frank Arute}, \bibinfo{person}{Kunal Arya},
  \bibinfo{person}{Ryan Babbush}, \bibinfo{person}{Dave Bacon},
  \bibinfo{person}{Joseph~C. Bardin}, \bibinfo{person}{Rami Barends},
  \bibinfo{person}{Sergio Boixo}, \bibinfo{person}{Michael Broughton},
  \bibinfo{person}{Bob~B. Buckley}, \bibinfo{person}{David~A. Buell},
  \bibinfo{person}{Brian Burkett}, \bibinfo{person}{Nicholas Bushnell},
  \bibinfo{person}{Yu Chen}, \bibinfo{person}{Zijun Chen}, \bibinfo{person}{Ben
  Chiaro}, \bibinfo{person}{Roberto Collins}, \bibinfo{person}{William
  Courtney}, \bibinfo{person}{Sean Demura}, \bibinfo{person}{Andrew Dunsworth},
  \bibinfo{person}{Edward Farhi}, \bibinfo{person}{Austin Fowler},
  \bibinfo{person}{Brooks Foxen}, \bibinfo{person}{Craig Gidney},
  \bibinfo{person}{Marissa Giustina}, \bibinfo{person}{Rob Graff},
  \bibinfo{person}{Steve Habegger}, \bibinfo{person}{Matthew~P. Harrigan},
  \bibinfo{person}{Alan Ho}, \bibinfo{person}{Sabrina Hong},
  \bibinfo{person}{Trent Huang}, \bibinfo{person}{L.~B. Ioffe},
  \bibinfo{person}{Sergei~V. Isakov}, \bibinfo{person}{Evan Jeffrey},
  \bibinfo{person}{Zhang Jiang}, \bibinfo{person}{Cody Jones},
  \bibinfo{person}{Dvir Kafri}, \bibinfo{person}{Kostyantyn Kechedzhi},
  \bibinfo{person}{Julian Kelly}, \bibinfo{person}{Seon Kim},
  \bibinfo{person}{Paul~V. Klimov}, \bibinfo{person}{Alexander~N. Korotkov},
  \bibinfo{person}{Fedor Kostritsa}, \bibinfo{person}{David Landhuis},
  \bibinfo{person}{Pavel Laptev}, \bibinfo{person}{Mike Lindmark},
  \bibinfo{person}{Martin Leib}, \bibinfo{person}{Erik Lucero},
  \bibinfo{person}{Orion Martin}, \bibinfo{person}{John~M. Martinis},
  \bibinfo{person}{Jarrod~R. McClean}, \bibinfo{person}{Matt McEwen},
  \bibinfo{person}{Anthony Megrant}, \bibinfo{person}{Xiao Mi},
  \bibinfo{person}{Masoud Mohseni}, \bibinfo{person}{Wojciech Mruczkiewicz},
  \bibinfo{person}{Josh Mutus}, \bibinfo{person}{Ofer Naaman},
  \bibinfo{person}{Matthew Neeley}, \bibinfo{person}{Charles Neill},
  \bibinfo{person}{Florian Neukart}, \bibinfo{person}{Hartmut Neven},
  \bibinfo{person}{Murphy~Yuezhen Niu}, \bibinfo{person}{Thomas~E. O'Brien},
  \bibinfo{person}{Bryan O'Gorman}, \bibinfo{person}{Eric Ostby},
  \bibinfo{person}{Andre Petukhov}, \bibinfo{person}{Harald Putterman},
  \bibinfo{person}{Chris Quintana}, \bibinfo{person}{Pedram Roushan},
  \bibinfo{person}{Nicholas~C. Rubin}, \bibinfo{person}{Daniel Sank},
  \bibinfo{person}{Kevin~J. Satzinger}, \bibinfo{person}{Andrea Skolik},
  \bibinfo{person}{Vadim Smelyanskiy}, \bibinfo{person}{Doug Strain},
  \bibinfo{person}{Michael Streif}, \bibinfo{person}{Kevin~J. Sung},
  \bibinfo{person}{Marco Szalay}, \bibinfo{person}{Amit Vainsencher},
  \bibinfo{person}{Theodore White}, \bibinfo{person}{Z.~Jamie Yao},
  \bibinfo{person}{Ping Yeh}, \bibinfo{person}{Adam Zalcman}, {and}
  \bibinfo{person}{Leo Zhou}.} \bibinfo{year}{2020}\natexlab{}.
\newblock \showarticletitle{Quantum {Approximate} {Optimization} of
  {Non}-{Planar} {Graph} {Problems} on a {Planar} {Superconducting}
  {Processor}}.
\newblock \bibinfo{journal}{\emph{arXiv:2004.04197 [quant-ph]}}
  (\bibinfo{date}{April} \bibinfo{year}{2020}).
\newblock
\urldef\tempurl%
\url{http://arxiv.org/abs/2004.04197}
\showURL{%
\tempurl}
\newblock
\shownote{arXiv: 2004.04197.}


\bibitem[\protect\citeauthoryear{Aspuru-Guzik, Dutoi, Love, and
  Head-Gordon}{Aspuru-Guzik et~al\mbox{.}}{2005}]%
        {science05-aspuru-guzik-dutoi-love-head-gordon-molecular-energies}
\bibfield{author}{\bibinfo{person}{Al{\'a}n Aspuru-Guzik},
  \bibinfo{person}{Anthony~D. Dutoi}, \bibinfo{person}{Peter~J. Love}, {and}
  \bibinfo{person}{Martin Head-Gordon}.} \bibinfo{year}{2005}\natexlab{}.
\newblock \showarticletitle{Simulated Quantum Computation of Molecular
  Energies}.
\newblock \bibinfo{journal}{\emph{Science}} \bibinfo{volume}{309},
  \bibinfo{number}{5741} (\bibinfo{year}{2005}), \bibinfo{pages}{1704--1707}.
\newblock
\showISSN{0036-8075}
\urldef\tempurl%
\url{https://doi.org/10.1126/science.1113479}
\showDOI{\tempurl}
\showeprint{https://science.sciencemag.org/content/309/5741/1704.full.pdf}


\bibitem[\protect\citeauthoryear{{Bhattacharjee}, {Saki}, {Alam},
  {Chattopadhyay}, and {Ghosh}}{{Bhattacharjee} et~al\mbox{.}}{2019}]%
        {iccad19-bhattacharjee-saki-alam-chattopadhyay-ghosh-muqut}
\bibfield{author}{\bibinfo{person}{D. {Bhattacharjee}}, \bibinfo{person}{A.~A.
  {Saki}}, \bibinfo{person}{M. {Alam}}, \bibinfo{person}{A. {Chattopadhyay}},
  {and} \bibinfo{person}{S. {Ghosh}}.} \bibinfo{year}{2019}\natexlab{}.
\newblock \showarticletitle{{MUQUT}: Multi-Constraint Quantum Circuit Mapping
  on {NISQ} Computers: Invited Paper}. In \bibinfo{booktitle}{\emph{2019
  IEEE/ACM International Conference on Computer-Aided Design (ICCAD)}}.
  \bibinfo{pages}{1--7}.
\newblock
\showISSN{1933-7760}
\urldef\tempurl%
\url{https://doi.org/10.1109/ICCAD45719.2019.8942132}
\showDOI{\tempurl}


\bibitem[\protect\citeauthoryear{Biamonte, Wittek, Pancotti, Rebentrost, Wiebe,
  and Lloyd}{Biamonte et~al\mbox{.}}{2017}]%
        {nature17-biamonte-wittek-pancotti-rebentrost-wiebe-lloyd-qml}
\bibfield{author}{\bibinfo{person}{Jacob Biamonte}, \bibinfo{person}{Peter
  Wittek}, \bibinfo{person}{Nicola Pancotti}, \bibinfo{person}{Patrick
  Rebentrost}, \bibinfo{person}{Nathan Wiebe}, {and} \bibinfo{person}{Seth
  Lloyd}.} \bibinfo{year}{2017}\natexlab{}.
\newblock \showarticletitle{Quantum machine learning}.
\newblock \bibinfo{journal}{\emph{Nature}} \bibinfo{volume}{549},
  \bibinfo{number}{7671} (\bibinfo{date}{Sept.} \bibinfo{year}{2017}),
  \bibinfo{pages}{195--202}.
\newblock
\showISSN{0028-0836, 1476-4687}
\urldef\tempurl%
\url{https://doi.org/10.1038/nature23474}
\showDOI{\tempurl}


\bibitem[\protect\citeauthoryear{Childs, Schoute, and Unsal}{Childs
  et~al\mbox{.}}{2019}]%
        {tqc19-childs-shoute-unsal-transformation}
\bibfield{author}{\bibinfo{person}{Andrew~M Childs}, \bibinfo{person}{Eddie
  Schoute}, {and} \bibinfo{person}{Cem~M Unsal}.}
  \bibinfo{year}{2019}\natexlab{}.
\newblock \showarticletitle{Circuit Transformations for Quantum Architectures}.
  In \bibinfo{booktitle}{\emph{14th Conference on the Theory of Quantum
  Computation, Communication and Cryptography (TQC 2019)}}. Schloss
  Dagstuhl-Leibniz-Zentrum fuer Informatik.
\newblock


\bibitem[\protect\citeauthoryear{de~Moura and Bj{\o}rner}{de~Moura and
  Bj{\o}rner}{2008}]%
        {book08-de-moura-bjorner-z3}
\bibfield{author}{\bibinfo{person}{Leonardo de Moura} {and}
  \bibinfo{person}{Nikolaj Bj{\o}rner}.} \bibinfo{year}{2008}\natexlab{}.
\newblock \showarticletitle{Z3: {An} {Efficient} {SMT} {Solver}}.
\newblock In \bibinfo{booktitle}{\emph{Tools and {Algorithms} for the
  {Construction} and {Analysis} of {Systems}}},
  \bibfield{editor}{\bibinfo{person}{David Hutchison}, \bibinfo{person}{Takeo
  Kanade}, \bibinfo{person}{Josef Kittler}, \bibinfo{person}{Jon~M. Kleinberg},
  \bibinfo{person}{Friedemann Mattern}, \bibinfo{person}{John~C. Mitchell},
  \bibinfo{person}{Moni Naor}, \bibinfo{person}{Oscar Nierstrasz},
  \bibinfo{person}{C.~Pandu~Rangan}, \bibinfo{person}{Bernhard Steffen},
  \bibinfo{person}{Madhu Sudan}, \bibinfo{person}{Demetri Terzopoulos},
  \bibinfo{person}{Doug Tygar}, \bibinfo{person}{Moshe~Y. Vardi},
  \bibinfo{person}{Gerhard Weikum}, \bibinfo{person}{C.~R. Ramakrishnan}, {and}
  \bibinfo{person}{Jakob Rehof}} (Eds.). Vol.~\bibinfo{volume}{4963}.
  \bibinfo{publisher}{Springer Berlin Heidelberg}, \bibinfo{address}{Berlin,
  Heidelberg}, \bibinfo{pages}{337--340}.
\newblock
\showISBNx{978-3-540-78799-0 978-3-540-78800-3}
\urldef\tempurl%
\url{https://doi.org/10.1007/978-3-540-78800-3_24}
\showDOI{\tempurl}
\newblock
\shownote{Series Title: Lecture Notes in Computer Science.}


\bibitem[\protect\citeauthoryear{Farhi, Goldstone, and Gutmann}{Farhi
  et~al\mbox{.}}{2014}]%
        {arxiv1411-farhi-goldstone-gutmann-qaoa}
\bibfield{author}{\bibinfo{person}{Edward Farhi}, \bibinfo{person}{Jeffrey
  Goldstone}, {and} \bibinfo{person}{Sam Gutmann}.}
  \bibinfo{year}{2014}\natexlab{}.
\newblock \bibinfo{title}{A Quantum Approximate Optimization Algorithm}.
\newblock
\newblock
\showeprint[arxiv]{1411.4028}~[quant-ph]


\bibitem[\protect\citeauthoryear{Gokhale, Javadi-Abhari, Earnest, Shi, and
  Chong}{Gokhale et~al\mbox{.}}{2020}]%
        {arxiv2004-gokhale-javadi-abhari-earnest-shi-chong-openpulse}
\bibfield{author}{\bibinfo{person}{Pranav Gokhale}, \bibinfo{person}{Ali
  Javadi-Abhari}, \bibinfo{person}{Nathan Earnest}, \bibinfo{person}{Yunong
  Shi}, {and} \bibinfo{person}{Frederic~T. Chong}.}
  \bibinfo{year}{2020}\natexlab{}.
\newblock \showarticletitle{Optimized {Quantum} {Compilation} for {Near}-{Term}
  {Algorithms} with {OpenPulse}}.
\newblock \bibinfo{journal}{\emph{arXiv:2004.11205 [quant-ph]}}
  (\bibinfo{date}{April} \bibinfo{year}{2020}).
\newblock
\urldef\tempurl%
\url{http://arxiv.org/abs/2004.11205}
\showURL{%
\tempurl}
\newblock
\shownote{arXiv: 2004.11205.}


\bibitem[\protect\citeauthoryear{Hadfield, Wang, O'Gorman, Rieffel, Venturelli,
  and Biswas}{Hadfield et~al\mbox{.}}{2019}]%
        {algorithms19-hadfield-wang-ogorman-rieffel-venturelli-biswas-qaoa}
\bibfield{author}{\bibinfo{person}{Stuart Hadfield}, \bibinfo{person}{Zhihui
  Wang}, \bibinfo{person}{Bryan O'Gorman}, \bibinfo{person}{Eleanor~G Rieffel},
  \bibinfo{person}{Davide Venturelli}, {and} \bibinfo{person}{Rupak Biswas}.}
  \bibinfo{year}{2019}\natexlab{}.
\newblock \showarticletitle{From the quantum approximate optimization algorithm
  to a quantum alternating operator ansatz}.
\newblock \bibinfo{journal}{\emph{Algorithms}} \bibinfo{volume}{12},
  \bibinfo{number}{2} (\bibinfo{year}{2019}), \bibinfo{pages}{34}.
\newblock


\bibitem[\protect\citeauthoryear{Li, Ding, and Xie}{Li et~al\mbox{.}}{2019}]%
        {asplos19-li-ding-xie-sabre}
\bibfield{author}{\bibinfo{person}{Gushu Li}, \bibinfo{person}{Yufei Ding},
  {and} \bibinfo{person}{Yuan Xie}.} \bibinfo{year}{2019}\natexlab{}.
\newblock \showarticletitle{Tackling the Qubit Mapping Problem for {NISQ}-Era
  Quantum Devices}. In \bibinfo{booktitle}{\emph{Proceedings of the
  Twenty-Fourth International Conference on Architectural Support for
  Programming Languages and Operating Systems}} (Providence, RI, USA)
  \emph{(\bibinfo{series}{ASPLOS '19})}. \bibinfo{publisher}{Association for
  Computing Machinery}, \bibinfo{address}{New York, NY, USA},
  \bibinfo{pages}{1001--1014}.
\newblock
\showISBNx{9781450362405}
\urldef\tempurl%
\url{https://doi.org/10.1145/3297858.3304023}
\showDOI{\tempurl}


\bibitem[\protect\citeauthoryear{{Maslov}, {Falconer}, and {Mosca}}{{Maslov}
  et~al\mbox{.}}{2008}]%
        {tcad08-maslov-falconer-mosca-placement}
\bibfield{author}{\bibinfo{person}{D. {Maslov}}, \bibinfo{person}{S.~M.
  {Falconer}}, {and} \bibinfo{person}{M. {Mosca}}.}
  \bibinfo{year}{2008}\natexlab{}.
\newblock \showarticletitle{Quantum Circuit Placement}.
\newblock \bibinfo{journal}{\emph{IEEE Transactions on Computer-Aided Design of
  Integrated Circuits and Systems}} \bibinfo{volume}{27}, \bibinfo{number}{4}
  (\bibinfo{date}{April} \bibinfo{year}{2008}), \bibinfo{pages}{752--763}.
\newblock
\showISSN{1937-4151}
\urldef\tempurl%
\url{https://doi.org/10.1109/TCAD.2008.917562}
\showDOI{\tempurl}


\bibitem[\protect\citeauthoryear{Murali, Linke, Martonosi, Javadi-Abhari,
  Nguyen, and Alderete}{Murali et~al\mbox{.}}{2019}]%
        {isca19-murali-linke-martonosi-abhari-nguyen-triq}
\bibfield{author}{\bibinfo{person}{Prakash Murali},
  \bibinfo{person}{Norbert~Matthias Linke}, \bibinfo{person}{Margaret
  Martonosi}, \bibinfo{person}{Ali Javadi-Abhari}, \bibinfo{person}{Nhung~Hong
  Nguyen}, {and} \bibinfo{person}{Cinthia~Huerta Alderete}.}
  \bibinfo{year}{2019}\natexlab{}.
\newblock \showarticletitle{Full-Stack, Real-System Quantum Computer Studies:
  Architectural Comparisons and Design Insights}. In
  \bibinfo{booktitle}{\emph{Proceedings of the 46th International Symposium on
  Computer Architecture}} (Phoenix, Arizona) \emph{(\bibinfo{series}{ISCA
  '19})}. \bibinfo{publisher}{Association for Computing Machinery},
  \bibinfo{address}{New York, NY, USA}, \bibinfo{pages}{527--540}.
\newblock
\showISBNx{9781450366694}
\urldef\tempurl%
\url{https://doi.org/10.1145/3307650.3322273}
\showDOI{\tempurl}


\bibitem[\protect\citeauthoryear{Nam, Ross, Su, Childs, and Maslov}{Nam
  et~al\mbox{.}}{2018}]%
        {npjqi18-nam-ross-su-childs-maslov-optimization}
\bibfield{author}{\bibinfo{person}{Yunseong Nam}, \bibinfo{person}{Neil~J.
  Ross}, \bibinfo{person}{Yuan Su}, \bibinfo{person}{Andrew~M. Childs}, {and}
  \bibinfo{person}{Dmitri Maslov}.} \bibinfo{year}{2018}\natexlab{}.
\newblock \showarticletitle{Automated optimization of large quantum circuits
  with continuous parameters}.
\newblock \bibinfo{journal}{\emph{npj Quantum Information}}
  \bibinfo{volume}{4}, \bibinfo{number}{1} (\bibinfo{date}{May}
  \bibinfo{year}{2018}).
\newblock
\showISSN{2056-6387}
\urldef\tempurl%
\url{https://doi.org/10.1038/s41534-018-0072-4}
\showDOI{\tempurl}


\bibitem[\protect\citeauthoryear{Nielsen and Chuang}{Nielsen and
  Chuang}{2010}]%
        {book10-nielsen-chuang}
\bibfield{author}{\bibinfo{person}{Michael~A Nielsen} {and}
  \bibinfo{person}{Isaac~L Chuang}.} \bibinfo{year}{2010}\natexlab{}.
\newblock \bibinfo{booktitle}{\emph{Quantum Computation and Quantum
  Information}}.
\newblock \bibinfo{publisher}{Cambridge University Press},
  \bibinfo{address}{Cambridge, UK}.
\newblock


\bibitem[\protect\citeauthoryear{Rieffel, Hadfield, Hogg, Mandr{\`a}, Marshall,
  Mossi, O'Gorman, Plamadeala, Tubman, Venturelli, Vinci, Wang, Wilson,
  Wudarski, and Biswas}{Rieffel et~al\mbox{.}}{2019}]%
        {arxiv1905-nasa-qaoa}
\bibfield{author}{\bibinfo{person}{Eleanor~G. Rieffel}, \bibinfo{person}{Stuart
  Hadfield}, \bibinfo{person}{Tad Hogg}, \bibinfo{person}{Salvatore
  Mandr{\`a}}, \bibinfo{person}{Jeffrey Marshall}, \bibinfo{person}{Gianni
  Mossi}, \bibinfo{person}{Bryan O'Gorman}, \bibinfo{person}{Eugeniu
  Plamadeala}, \bibinfo{person}{Norm~M. Tubman}, \bibinfo{person}{Davide
  Venturelli}, \bibinfo{person}{Walter Vinci}, \bibinfo{person}{Zhihui Wang},
  \bibinfo{person}{Max Wilson}, \bibinfo{person}{Filip Wudarski}, {and}
  \bibinfo{person}{Rupak Biswas}.} \bibinfo{year}{2019}\natexlab{}.
\newblock \showarticletitle{From {Ans}{\textbackslash}"atze to {Z}-gates: a
  {NASA} {View} of {Quantum} {Computing}}.
\newblock \bibinfo{journal}{\emph{arXiv:1905.02860 [quant-ph]}}
  (\bibinfo{date}{May} \bibinfo{year}{2019}).
\newblock
\urldef\tempurl%
\url{http://arxiv.org/abs/1905.02860}
\showURL{%
\tempurl}
\newblock
\shownote{arXiv: 1905.02860.}


\bibitem[\protect\citeauthoryear{Shafaei, Saeedi, and Pedram}{Shafaei
  et~al\mbox{.}}{2013}]%
        {dac13-shafaei-saeedi-pedram-linear}
\bibfield{author}{\bibinfo{person}{Alireza Shafaei}, \bibinfo{person}{Mehdi
  Saeedi}, {and} \bibinfo{person}{Massoud Pedram}.}
  \bibinfo{year}{2013}\natexlab{}.
\newblock \showarticletitle{Optimization of Quantum Circuits for Interaction
  Distance in Linear Nearest Neighbor Architectures}. In
  \bibinfo{booktitle}{\emph{Proceedings of the 50th Annual Design Automation
  Conference}} (Austin, Texas) \emph{(\bibinfo{series}{DAC '13})}.
  \bibinfo{publisher}{Association for Computing Machinery},
  \bibinfo{address}{New York, NY, USA}, Article \bibinfo{articleno}{41},
  \bibinfo{numpages}{6}~pages.
\newblock
\showISBNx{9781450320719}
\urldef\tempurl%
\url{https://doi.org/10.1145/2463209.2488785}
\showDOI{\tempurl}


\bibitem[\protect\citeauthoryear{{Shafaei}, {Saeedi}, and {Pedram}}{{Shafaei}
  et~al\mbox{.}}{2014}]%
        {aspdac14-shafaei-saeedi-pedram-2014-placement-2d}
\bibfield{author}{\bibinfo{person}{A. {Shafaei}}, \bibinfo{person}{M.
  {Saeedi}}, {and} \bibinfo{person}{M. {Pedram}}.}
  \bibinfo{year}{2014}\natexlab{}.
\newblock \showarticletitle{Qubit placement to minimize communication overhead
  in 2{D} quantum architectures}. In \bibinfo{booktitle}{\emph{2014 19th Asia
  and South Pacific Design Automation Conference (ASP-DAC)}}.
  \bibinfo{pages}{495--500}.
\newblock
\showISSN{2153-697X}
\urldef\tempurl%
\url{https://doi.org/10.1109/ASPDAC.2014.6742940}
\showDOI{\tempurl}


\bibitem[\protect\citeauthoryear{Shor}{Shor}{1994}]%
        {sfcs94-shor-algorithms}
\bibfield{author}{\bibinfo{person}{P.W. Shor}.}
  \bibinfo{year}{1994}\natexlab{}.
\newblock \showarticletitle{Algorithms for quantum computation: discrete
  logarithms and factoring}. In \bibinfo{booktitle}{\emph{Proceedings 35th
  {Annual} {Symposium} on {Foundations} of {Computer} {Science}}}.
  \bibinfo{publisher}{IEEE Comput. Soc. Press}, \bibinfo{address}{Santa Fe, NM,
  USA}, \bibinfo{pages}{124--134}.
\newblock
\showISBNx{978-0-8186-6580-6}
\urldef\tempurl%
\url{https://doi.org/10.1109/SFCS.1994.365700}
\showDOI{\tempurl}


\bibitem[\protect\citeauthoryear{Siraichi, Santos, Collange, and
  Pereira}{Siraichi et~al\mbox{.}}{2018}]%
        {cgo18-siraichi-santos-collange-pereira-allocation}
\bibfield{author}{\bibinfo{person}{Marcos~Yukio Siraichi},
  \bibinfo{person}{Vin\'{\i}cius Fernandes~Dos Santos},
  \bibinfo{person}{Sylvain Collange}, {and} \bibinfo{person}{Fernando
  Magno~Quintao Pereira}.} \bibinfo{year}{2018}\natexlab{}.
\newblock \showarticletitle{Qubit Allocation}. In
  \bibinfo{booktitle}{\emph{Proceedings of the 2018 International Symposium on
  Code Generation and Optimization}} (Vienna, Austria)
  \emph{(\bibinfo{series}{CGO 2018})}. \bibinfo{publisher}{Association for
  Computing Machinery}, \bibinfo{address}{New York, NY, USA},
  \bibinfo{pages}{113--125}.
\newblock
\showISBNx{9781450356176}
\urldef\tempurl%
\url{https://doi.org/10.1145/3168822}
\showDOI{\tempurl}


\bibitem[\protect\citeauthoryear{Sivarajah, Dilkes, Cowtan, Simmons, Edgington,
  and Duncan}{Sivarajah et~al\mbox{.}}{2020}]%
        {arxiv2003-sivarajah-dilkes-cowtan-simmons-edgington-duncan-tket}
\bibfield{author}{\bibinfo{person}{Seyon Sivarajah}, \bibinfo{person}{Silas
  Dilkes}, \bibinfo{person}{Alexander Cowtan}, \bibinfo{person}{Will Simmons},
  \bibinfo{person}{Alec Edgington}, {and} \bibinfo{person}{Ross Duncan}.}
  \bibinfo{year}{2020}\natexlab{}.
\newblock \showarticletitle{{t|ket>} : {A} {Retargetable} {Compiler} for {NISQ}
  {Devices}}.
\newblock \bibinfo{journal}{\emph{arXiv:2003.10611 [quant-ph]}}
  (\bibinfo{date}{March} \bibinfo{year}{2020}).
\newblock
\urldef\tempurl%
\url{http://arxiv.org/abs/2003.10611}
\showURL{%
\tempurl}
\newblock
\shownote{arXiv: 2003.10611.}


\bibitem[\protect\citeauthoryear{{Tan} and {Cong}}{{Tan} and {Cong}}{2020}]%
        {tc20-tan-cong-optimality}
\bibfield{author}{\bibinfo{person}{B. {Tan}} {and} \bibinfo{person}{J.
  {Cong}}.} \bibinfo{year}{2020}\natexlab{}.
\newblock \showarticletitle{Optimality Study of Existing Quantum Computing
  Layout Synthesis Tools}.
\newblock \bibinfo{journal}{\emph{IEEE Trans. Comput.}} (\bibinfo{year}{2020}).
\newblock
\showeprint[arxiv]{2002.09783}


\bibitem[\protect\citeauthoryear{Tannu and Qureshi}{Tannu and Qureshi}{2019}]%
        {asplos19-tannu-qureshi-variability}
\bibfield{author}{\bibinfo{person}{Swamit~S. Tannu} {and}
  \bibinfo{person}{Moinuddin~K. Qureshi}.} \bibinfo{year}{2019}\natexlab{}.
\newblock \showarticletitle{Not All Qubits Are Created Equal: A Case for
  Variability-Aware Policies for {NISQ}-Era Quantum Computers}. In
  \bibinfo{booktitle}{\emph{Proceedings of the Twenty-Fourth International
  Conference on Architectural Support for Programming Languages and Operating
  Systems}} (Providence, RI, USA) \emph{(\bibinfo{series}{ASPLOS '19})}.
  \bibinfo{publisher}{Association for Computing Machinery},
  \bibinfo{address}{New York, NY, USA}, \bibinfo{pages}{987--999}.
\newblock
\showISBNx{9781450362405}
\urldef\tempurl%
\url{https://doi.org/10.1145/3297858.3304007}
\showDOI{\tempurl}


\bibitem[\protect\citeauthoryear{Wille, Burgholzer, and Zulehner}{Wille
  et~al\mbox{.}}{2019}]%
        {dac19-wille-burgholzer-zulehner-minimal}
\bibfield{author}{\bibinfo{person}{Robert Wille}, \bibinfo{person}{Lukas
  Burgholzer}, {and} \bibinfo{person}{Alwin Zulehner}.}
  \bibinfo{year}{2019}\natexlab{}.
\newblock \showarticletitle{Mapping Quantum Circuits to {IBM} {QX}
  Architectures Using the Minimal Number of {SWAP} and {H} Operations}. In
  \bibinfo{booktitle}{\emph{Proceedings of the 56th Annual Design Automation
  Conference 2019}} (Las Vegas, NV, USA) \emph{(\bibinfo{series}{DAC '19})}.
  \bibinfo{publisher}{Association for Computing Machinery},
  \bibinfo{address}{New York, NY, USA}, Article \bibinfo{articleno}{142},
  \bibinfo{numpages}{6}~pages.
\newblock
\showISBNx{9781450367257}
\urldef\tempurl%
\url{https://doi.org/10.1145/3316781.3317859}
\showDOI{\tempurl}


\bibitem[\protect\citeauthoryear{{Wille}, {Gro{\ss}e}, {Teuber}, {Dueck}, and
  {Drechsler}}{{Wille} et~al\mbox{.}}{2008}]%
        {ismvl08-wille-grobe-teuber-dueck-dreschsler-revlib}
\bibfield{author}{\bibinfo{person}{R. {Wille}}, \bibinfo{person}{D.
  {Gro{\ss}e}}, \bibinfo{person}{L. {Teuber}}, \bibinfo{person}{G.~W. {Dueck}},
  {and} \bibinfo{person}{R. {Drechsler}}.} \bibinfo{year}{2008}\natexlab{}.
\newblock \showarticletitle{{RevLib}: An Online Resource for Reversible
  Functions and Reversible Circuits}. In \bibinfo{booktitle}{\emph{38th
  International Symposium on Multiple Valued Logic (ismvl 2008)}}.
  \bibinfo{pages}{220--225}.
\newblock
\showISSN{2378-2226}
\urldef\tempurl%
\url{https://doi.org/10.1109/ISMVL.2008.43}
\showDOI{\tempurl}


\bibitem[\protect\citeauthoryear{{Wille}, {Lye}, and {Drechsler}}{{Wille}
  et~al\mbox{.}}{2014}]%
        {aspdac14-wille-lye-dreschsler-optimal}
\bibfield{author}{\bibinfo{person}{R. {Wille}}, \bibinfo{person}{A. {Lye}},
  {and} \bibinfo{person}{R. {Drechsler}}.} \bibinfo{year}{2014}\natexlab{}.
\newblock \showarticletitle{Optimal {SWAP} gate insertion for nearest neighbor
  quantum circuits}. In \bibinfo{booktitle}{\emph{2014 19th Asia and South
  Pacific Design Automation Conference (ASP-DAC)}}. \bibinfo{pages}{489--494}.
\newblock
\showISSN{2153-697X}
\urldef\tempurl%
\url{https://doi.org/10.1109/ASPDAC.2014.6742939}
\showDOI{\tempurl}


\bibitem[\protect\citeauthoryear{{Zulehner}, {Paler}, and {Wille}}{{Zulehner}
  et~al\mbox{.}}{2018}]%
        {date18-zulehner-paler-wille-efficient}
\bibfield{author}{\bibinfo{person}{A. {Zulehner}}, \bibinfo{person}{A.
  {Paler}}, {and} \bibinfo{person}{R. {Wille}}.}
  \bibinfo{year}{2018}\natexlab{}.
\newblock \showarticletitle{Efficient mapping of quantum circuits to the {IBM}
  {QX} architectures}. In \bibinfo{booktitle}{\emph{2018 Design, Automation
  Test in Europe Conference Exhibition (DATE)}}. \bibinfo{pages}{1135--1138}.
\newblock
\showISSN{1558-1101}
\urldef\tempurl%
\url{https://doi.org/10.23919/DATE.2018.8342181}
\showDOI{\tempurl}


\end{thebibliography}

\clearpage
\small

\begin{table*}[htbp]
    \caption{Evaluation of OLSQ}
    \vspace{-1\baselineskip}
    \label{tab:OLSQ}
    \begin{minipage}{\textwidth}
        \begin{center}
            \begin{tabular}{lllllllllllllllllll}
            \toprule
            &&&\multicolumn{4}{l}{Wille et al., 2019} &\multicolumn{4}{l}{OLSQ-SWAP} &\multicolumn{4}{l}{OLSQ-Depth} &\multicolumn{4}{l}{OLSQ-Fidelity} \\
            Program &$M$ &Architecture &c &d &f &t &c &d &f &t &c &d &f &t &c &d &f &t \\
            \midrule
            or &3 &IBM QX2 &0 &9 &0.594 &0.2 &0 &9 &0.589 &5 &0 &9 &0.625 &4 &0 &9 &0.625 &5 \\
            adder &4 &IBM QX2 &6 &18 &0.335 &1 &3 &16 &0.407 &40 &3 &16 &0.391 &40 &3 &16 &0.431 &2E3 \\
            adder &4 &Grid2by3* &0 &12 &0.462 &2 &0 &12 &0.462 &10 &0 &12 &0.462 &10 &0 &12 &0.462 &10 \\
            adder &4 &Grid2by4* &0 &12 &0.462 &1E3 &0 &12 &0.462 &10 &0 &12 &0.462 &10 &0 &12 &0.462 &20 \\
            qaoa5 &5 &IBM QX2 &0 &15 &0.470 &0.3 &0 &15 &0.466 &10 &3 &15 &0.389 &10 &0 &15 &0.476 &10 \\
            mod5mils\_65 &5 &IBM QX2 &6 &28 &0.290 &2 &6 &28 &0.299 &1E2 &12 &25 &0.198 &90 &6 &28 &0.301 &7E2 \\
            4gt\_13\_92 &5 &IBM QX2 &0 &39 &0.160 &2 &0 &39 &0.171 & 2E2 &12 &39 &0.107 &1E2 &0 &39 &0.171 &3E2 \\
            4mod5-v1\_22 &5 &IBM QX2 &3 &16 &0.313 &0.5 &3 &16 &0.391 &20 &6 &16 &0.352 &20 &3 &16 &0.391 &30 \\
            4mod5-v1\_22 &5 &Grid2by3* &9 &22 &0.271 &8 &9 &21 &0.271 &4E2 &12 &21 &0.236 &1E2 &9 &22 &0.271 &3E3 \\
            4mod5-v1\_22 &5 &Grid2by4* &9 &18 &0.271 &1E4 &9 &22 &0.271 &7E2 &9 &21 &0.206 &2E2 &9 &24 &0.271 &8E2 \\
            queko\_05\_0 &16 &Aspen-4* &\multicolumn{4}{c}{out of memory (32GB)} &0 &6 &0.148 &70 &0 &6 &0.148 &70 &0 &6 &0.148 &9E2 \\
            queko\_10\_3 &16 &Aspen-4* &\multicolumn{4}{c}{out of memory (32GB)} &0 &11 &0.076 &8E2 &0 &11 &0.076 &8E2 &0 &11 &0.076 &4E3 \\
            queko\_15\_1 &16 &Aspen-4* &\multicolumn{4}{c}{out of memory (32GB)} &0 &16 &0.037 &5E3 &0 &16 &0.037 &5E3 &0 &16 &0.037 &1E4 \\
            \bottomrule
            \end{tabular}
        \end{center}
        \footnotesize
        \emph{Note:} $M$ is logical qubit count of the program; c is additional $CX$ count; d is depth; f is fidelity (up to three digit precision); t is compilation time (up to one significant digit precision, `E' means 10 exponential). We do not have fidelity profile of the architectures with *, so we use a uniform fidelity profile based on the average fidelity of IBM QX2.
    \end{minipage}

\end{table*}

\begin{table*}[htbp]
    \setlength{\tabcolsep}{4pt}
    \caption{Evaluation of TB-OLSQ}
    \vspace{-1\baselineskip}
    \label{tab:tb-OLSQ}
    \begin{minipage}{\textwidth}
        \begin{center}
            \begin{tabular}{lllllllllllllllll}
            \toprule
            &&&\multicolumn{3}{l}{$\text{t}|\text{ket}\rangle$} &\multicolumn{3}{l}{TB-OLSQ-SWAP} &\multirow{2}{4em}{$CX$ Cost Reduction} &\multicolumn{3}{l}{TriQ} &\multicolumn{3}{l}{TB-OLSQ-Fidelity} &\multirow{2}{2.8em}{Fidelity Boost}\\
            Program &$M$ &Architecture &c &d &f &c &d &f & &c &d &f &c &d &f \\
            \midrule
            or &3 &IBM QX2 &0 &9 &0.625 &0 &9 &0.625 &0 &0 &9 &0.626 &0 &9 &0.625 &1.00x \\
            adder &4 &IBM QX2 &12 &27 &0.246 &3 &16 &0.439 &75\% &6 &24 &0.371 &3 &16 &0.431 &1.16x \\
            qaoa5 &5 &IBM QX2 &3 &17 &0.396 &0 &15 &0.467 &100\% &0 &16 &0.475 &0 &15 &0.476 &1.00x \\
            mod5mils\_65 &5 &IBM QX2 &12 &34 &0.203 &6 &29 &0.249 &50\% &12 &50 &0.230 &6 &27 &0.302 &1.31x \\
            4gt13\_92 &5 &IBM QX2 &21 &64 &5.72E-2 &0 &39 &0.155 &100\% &0 &48 &0.165 &0 &39 &0.171 &1.04x \\
            4mod5-v1\_22 &5 &IBM QX2 &12 &29 &0.282 &3 &16 &0.384 &75\% &3 &24 &0.406 &3 &16 &0.391 &0.96x\\
            queko\_05\_0 &16 &Aspen-4 &3 &10 &0.129 &0 &6 &0.148 &100\% &0 &6 &0.148 &0 &6 &0.148 &1.00x \\
            queko\_10\_3 &16 &Aspen-4 &45 &45 &9.59E-3 &0 &11 &0.076 &100\% &0 &11 &0.076 &0 &15 &0.074 &1.03x \\
            queko\_15\_1 &16 &Aspen-4 &114 &58 &1.96E-4 &0 &16 &3.74E-2 &100\% &0 &16 &3.74E-2 &0 &19 &3.62E-2 &1.03x \\
            or &3 &Melbourne &6 &18 &0.350 &6 &16 &0.276 &0 &6 &23 &0.364 &6 &17 &0.350 &0.96x \\
            adder &4 &Melbourne &3 &14 &0.216 &0 &12 &0.363 &100\% &0 &14 &0.365 &0 &12 &0.369 &1.01x \\
            qaoa5 &5 &Melbourne &0 &15 &0.340 &0 &15 &0.114 &0 &0 &17 &0.381 &0 &15 &0.424 &1.11x \\
            mod5mils\_65 &5 &Melbourne &18 &45 &0.118 &18 &34 &3.32E-2 &0 &21 &65 &8.88E-2 &18 &43 &0.103 &1.16x \\
            4gt13\_92 &5 &Melbourne &42 &84 &2.11E-3 &30 &68 &5.36E-6 &28.6\% &39 &116 &1.48E-2 &36 &72 &1.74E-2 &1.18x \\
            4mod5-v1\_22 &5 &Melbourne &9 &24 &1.83E-4 &9 &16 &0.131 &0 &9 &38 &0.215 &9 &25 &0.247 &1.15x\\
            tof\_4 &7 &Melbourne &3 &53 &2.04E-3 &3 &47 &2.14E-3 &0 &6 &50 &9.57E-2 &3 &47 &1.22E-1 &1.27x \\
            barenco\_tof\_4 &7 &Melbourne &27 &91 &3.93E-4 &15 &75 &5.77E-4 &44.4\% &21 &100 &2.22E-2 &24 &78 &2.18E-2 &0.98x \\
            tof\_5 &9 &Melbourne &21 &68 &3.09E-4 &3 &62 &1.00E-3 & 85.7\% &6 &63 &2.85E-2 &3 &62 &4.63E-2 &1.62x \\
            barenco\_tof\_5 &9 &Melbourne &51 &146 &3.02E-10 &18 &81 &6.37E-6 &64.7\% &39 &147 &1.39E-3 &39 &93 &1.83E-3 &1.32x \\
            mod\_mult\_55 &9 &Melbourne &36 &78 &1.60E-6 &24 &71 &3.18E-5 &33.3\% &50 &126 &7.76E-4 &24 &68 &7.77E-3 &10.0x \\
            vbe\_adder\_3 &10 &Melbourne &48 &96 &2.80E-8 &24 &59 &1.00E-8 &50.0\% &45 &124 &1.53E-3 &27 &61 &4.99E-3 &3.26x\\
            rc\_adder\_6 &14 &Melbourne &174 &186 &5.44E-22 &27 &80 &1.89E-7 &84.5\% &\multicolumn{3}{c}{timeout (24 hrs)} &27 &85 &2.01E-6 \\
            \multicolumn{3}{l}{\textbf{Geometric Mean}} &&&&&& &69.2\% &&&&&& &1.30x \\
            \bottomrule
            \end{tabular}
        \end{center}
        \footnotesize
        \emph{Note:} $M$ is logical qubit count of the program; c is additional $CX$ count; d is depth; f is fidelity (up to three significant digit precision). $CX$ cost reduction is the difference of c of TB-OLSQ-SWAP and $\text{t}|\text{ket}\rangle$ normalized by the c of $\text{t}|\text{ket}\rangle$. Fidelity boost is the ratio of the f of TB-OLSQ-Fidelity and TriQ. We do not have fidelity profile of the architectures with *, so we use a uniform fidelity profile based on the average fidelity of IBM QX2.
    \end{minipage}

\end{table*}

\begin{table*}[htbp]
    \caption{Evaluation of QAOA-OLSQ}
    \vspace{-1\baselineskip}
    \label{tab:qaoa-OLSQ}
    \begin{minipage}{\textwidth}
        \begin{center}
            \begin{tabular}{lllllllllll}
                \toprule
                &\multicolumn{2}{l}{$\text{t}|\text{ket}\rangle$} &\multicolumn{2}{l}{TB-OLSQ} &\multirow{2}{4em}{Depth Reduction} &\multirow{2}{4em}{SWAP Reduction} &\multicolumn{2}{l}{QAOA-OLSQ} &\multirow{2}{4em}{Depth Reduction} &\multirow{2}{4em}{SWAP Reduction} \\
                $M$ &Depth &SWAP &Depth &SWAP & & &Depth &SWAP &  & \\
                \midrule
                10 &16 &7.3 &6.9 &7.3 &56.7\% &0 &6.5 &5.5 &59.3\% &23.6\% \\
                12 &17.8 &11.7 &8.5 &9.3 &52.3\% &20.4\% &5.6 &5.8 &67.3\% &46.2\% \\
                14 &19.0 &13.2 &9.0 &12.3 &52.6\% &6.8\% &6.0 &6.6 &68.3\% &48.0\% \\
                16 &21.7 &20.2 &9.1 &13.6 &58.2\% &32.7\% &6.4 &6.9 &70.2\% &62.6\% \\
                18 &25.5 &26.7 &8.9 &14.5 &64.9\% &45.7\% &6.0 &8.3 &75.5\% &65.7\% \\
                20 &30.6 &37.5 &9.3 &16.3 &68.9\% &57.7\% &7.2 &10.8 &75.7\% &68.8\% \\
                22 &29.8 &38.4 &10.3 &17.8 &65.4\% &53.6\% &7.8 &14.2 &73.7\% &61.8\% \\
                \multicolumn{3}{l}{\textbf{Geometric Mean}}     &     &     &59.5\% &29.4\%  &    &     &70.2\% &53.8\% \\
                \bottomrule
            \end{tabular}
        \end{center}
        \footnotesize
        \emph{Note:} $M$ is the node count of the 3-regular graphs and also the logic qubit count. All the $ZZ$-phase gates and SWAP gates are seen to have unity depth. Depth and SWAP reduction is the difference between QAOA-OLSQ and $\text{t}|\text{ket}\rangle$ normalized by $\text{t}|\text{ket}\rangle$ result. All data are geometrical means and have one digit precision.
    \end{minipage}
\end{table*}

\end{document}